\documentclass[aps,prb,twocolumn,superscriptaddress,nofootinbib,longbibliography]{revtex4-1}
\usepackage{amssymb,graphicx,color}
\usepackage[intlimits]{amsmath}
\usepackage[english]{babel}
\usepackage[colorlinks]{hyperref}
\usepackage{comment}
\def\beq{\begin{equation}}
\def\eeq{\end{equation}}
\def\bea{\begin{eqnarray}}
\def\eea{\end{eqnarray}}
\def\barr{\begin{array}}
\def\earr{\end{array}}

\begin{document}

\title{Enhancement of crossed Andreev reflection in a Kitaev ladder connected to normal metal leads}
 \author{Ritu Nehra}
 \author{Devendra Singh Bhakuni}
 \author{Auditya Sharma} 
 \affiliation{ Department of Physics, Indian Institute of Science Education and Research, Bhopal 462066, India.}
 \author{Abhiram Soori}
\email{abhirams@uohyd.ac.in}
 \affiliation{ Department of Physics, Indian Institute of Science Education and Research, Bhopal 462066, India.}
\affiliation{ School of Physics, University of Hyderabad, C.R. Rao Road, Gachibowli, Hyderabad 500046,  India.}

\begin{abstract} 
We study nonlocal transport in a two-leg Kitaev ladder connected to
two normal metals. The coupling between the two legs of the ladder
when the legs are maintained at a (large) superconducting phase
difference, results in the creation of subgap Andreev states. These
states in turn are responsible for the enhancement of crossed Andreev
reflection. We find that tuning the different parameters of the system
suitably leads to enhancement of crossed Andreev reflection signalled
by transconductance acquiring the most negative value
possible. Furthermore, subgap states cause oscillations of the
transconductance as a function of various system parameters such as
chemical potential and ladder length, which are seen to be a
consequence of Fabry-P\'erot resonance.
\end{abstract}
%\pacs{}
\maketitle

\section{Introduction}
In a superconductor (S), there is a quasiparticle energy gap at the
Fermi energy which curbs the flow of quasiparticles into the
superconductor at low bias. However, there is a Cooper pair condensate
which can absorb the current injected from a normal metal~(N) lead and
this happens by the phenomenon of Andreev reflection~(AR).  The subgap
electron (with energy $E>0$) in the normal metal pairs up with another
electron below Fermi energy (with energy $-E$) and forms a Cooper pair
in the superconducting region. This phenomenon was first studied by
Andreev~\cite{andr64} and since then, it has been extensively studied
theoretically and experimentally for several decades in various condensed
matter systems~\cite{btk,kasta,kseng01,lutchyn2010majorana,oreg2010helical,
stanescu2011majorana,mourik2012signatures,albrecht2016exponential,
beiranvand2016tunable,tao2018superconductivity,wang2018andreev,lv2018magnetoanisotropic}.  Over the years,
Andreev reflection has been employed as a tool in a wide variety of
problems ranging from distinguishing between singlet and triplet
states~\cite{kseng01} to topological phase
transitions~\cite{lutchyn2010majorana} to experimental
signatures~\cite{mourik2012signatures,albrecht2016exponential} of
Majorana fermions~\cite{kitaev2001unpaired}. Also, intriguing
transport properties of topological
superconductors~\cite{qi2011topological,li2018selective} and junctions of
superconductors with topological insulators~\cite{soori2013transport}
have been understood in terms of Andreev reflection.  

Crossed Andreev reflection~(CAR)
is a variant of Andreev reflection which happens in a system consisting of two normal metallic
leads attached to a
superconductor~\cite{melin2009r,beckmann2004evidence,beckmann2006d,melin2004sign,bovzovic2002coherent,
  dong2003coherent,yamashita2003crossed,reinthaler2013proposal,yeyati2007entangled,he2014correlated,
  gomez2012selective,linder2014superconducting,linder2009spin,wang2015quantized,crepin2013even,
  chen2015long,chtchelkatchev2003superconducting,russo2005experimental,byers1995probing,
  deutscher2000coupling,falci2001correlated,zhang2018spin,islam17,beiranvand2017nonlocal}.  
  In this process, an
electron incident on the superconductor from the first normal metal ($N_1$) results
in a hole in the second normal metal ($N_2$), injecting a Cooper pair
into the superconductor.
% This happens when the distance between the two N's connected to the S is
% less than the superconducting coherence length.
% Also, CAR contributes to the rich pattern of conductances in one-dimensional
% N-S-N systems with $p-$wave S that hosts Majorana fermions at its 
% ends~\cite{thaku15}.
However, the electron incident from $N_1$ also results in an electron transmitted~(ET)
into $N_2$ and this process contributes a current which is opposite in sign to that of CAR.
A negative differential transconductance between $N_1$ and $N_2$ is strong evidence 
of CAR. But typically ET dominates CAR and a negative
transconductance for a given set of parameters is extremely rare~\cite{melin2009r,
chtchelkatchev2003superconducting,zhang2018electrically,zhang2018spin,soori2017enhancement}. 

Ladder systems have proven to be a rich playground for the exploration
of physics in a variety of contexts
~\cite{nehra2018many,hugel2014chiral,sil2008metal,sun2016topological,wang2016flux,
wu2012topological,wakatsuki2014majorana,Padavic2018topological,soori2017enhancement}.
In a recent piece of work~\cite{soori2017enhancement}, it was shown that when a
superconducting ladder is sandwiched between two normal metal leads,
the CAR can be enhanced by tuning the system parameters appropriately. When the phase
difference between the two legs of the ladder is fixed at $\pi$ and
for a strong enough coupling between the two legs (`strong' compared
to the superconducting gap in the individual leg of the ladder), it
was shown that the transconductance can be varied across a range of
values from one extreme ($-2e^2/h$) to the other
($+2e^2/h$).  The key reason why the ladder geometry
proves useful is that with a suitable phase difference between the
chains and by tuning the coupling between the chains, subgap states
which are responsible for enhancement of CAR can be created.  In
this paper, we address the question of whether a ladder made out of
Kitaev chains connected to two normal metallic leads can result in
enhanced CAR, and answer in the affirmative.

The key findings of this paper are as follows. Subgap Andreev states 
arise when a nonzero phase difference is maintained between the two legs of the ladder 
accompanied by a finite inter-leg hopping. The gap closes when the phase difference 
is set to $\pi$ and the inter-leg hopping crosses a critical value which is 
determined by the chemical potential in contrast to the spinful electronic model studied
earlier~\cite{soori2017enhancement}. The appearance of these subgap
Andreev states provides the propagating modes which in turn enhance
both CAR and ET. We find that by choosing parameters appropriately it
is possible to enhance CAR (and ET for different parameters) to
its highest possible value. Varying the ladder
length also provides very rich behavior. For small system sizes a
modest enhancement is observed even below the critical value of the
inter-leg hopping while beyond the critical value, CAR can be seen to
touch its extreme value. Below the critical value of the inter-leg hopping, the
transport is suppressed for larger lengths due to the presence of
decaying modes and above the critical value an oscillatory behavior is
observed where CAR and ET dominate alternatively. This is seen to be a
consequence of Fabry-P\'erot resonance~\cite{soori2012,soori2017enhancement}.

\begin{figure*}
\includegraphics[scale=1.1]{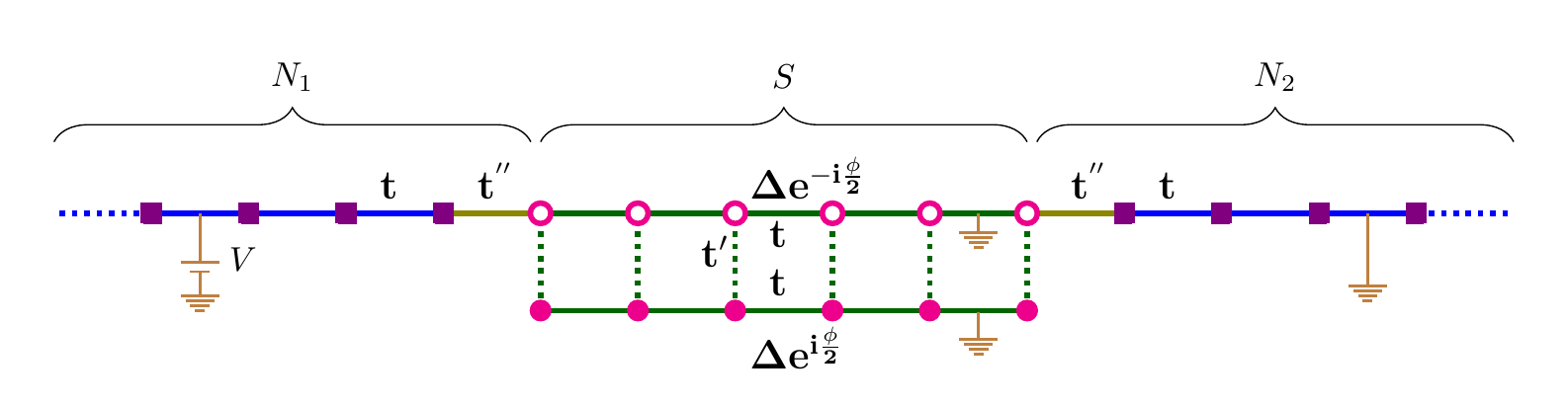}
\caption{Schematic diagram of a superconducting ladder~(S) connected to two normal metals~($N_1$,$N_2$).
The left metal($N_1$) is maintained at a bias voltage $V$ while the superconducting ladder~(S)
and the right metal~($N_2$) are grounded. The intra-leg hopping of the Kitaev ladder and the 
hopping in the normal metals are $t$, the inter-leg ladder hopping is $t^{\prime}$ with chemical potential
$\mu$ in all three regions. There is a superconducting pairing term $\Delta$ in each leg of
the ladder with opposite superconducting phase factors~($\mp\frac{\phi}{2}$). The normal metals are connected
to the upper leg of the ladder with a hopping $t^{\prime\prime}$.}
\label{fig:1}
\end{figure*}

The organization of our paper is as follows. The next section starts
with a description of the model Hamiltonian and proceeds to a
discussion of subgap Andreev states. The following section is about
the boundary conditions, wavefunctions, scattering amplitudes and the
calculation of transconductance. A Results and Analysis section puts
together all the findings. The following section explores
potential experimental realization of some of these phenomena.  This
is followed by a concluding summary section.

\section{Model Hamiltonian}
The system under study consists of two normal metal leads coupled to a superconducting ladder 
as shown in Fig.~\ref{fig:1}. The superconducting ladder is made out of two Kitaev chains maintained 
at a phase difference of $\phi$. The first chain has the superconducting pair potential $\Delta e^{-i\phi/2}$,
while the second chain has $\Delta e^{+i\phi/2}$. The two chains 
are connected by a hopping $t'$ at each site as shown. The hopping in normal metal leads and the two 
Kitaev chains is $t$. The normal metal leads are connected to the upper Kitaev chain by a hopping $t''$.

The Hamiltonian for the metallic regions is given by
\begin{align}
\label{eq:1}H_{N_1}=-t\displaystyle\sum_{n\le -1}(c_{n+1}^{\dagger}c_{n}+c_{n}^{\dagger}c_{n+1})
-\mu\displaystyle\sum_{n\le 0}c_{n}^{\dagger}c_{n}\\
\label{eq:2} H_{N_2}=-t\displaystyle\sum_{n\ge L+1}(c_{n+1}^{\dagger}c_{n}+c_{n}^{\dagger}c_{n+1})
-\mu\displaystyle\sum_{n\ge L+1}c_{n}^{\dagger}c_{n},
\end{align}
where $t$ is the hopping amplitude, the $c^{\dagger}_n(c_n)$ are
creation (annihilation) operators on the normal metals~($N_1$
for $n\le 0$ or $N_2$ for $n\ge L+1$) and $\mu$ is the chemical
potential. The dispersion in the normal metallic regions is the standard $E=\mp(2t\cos{ka}+\mu)$, 
where $-$~($+$) sign corresponds
to electrons (holes). The Hamiltonian for the Kitaev ladder is
given by
\begin{align}
\label{eq:3}
H_{S}=&\displaystyle\sum_{1\le n\le L-1\atop \sigma=1,2}-[(t c_{n+1,\sigma}^{\dagger}c_{n,\sigma}
+\Delta e^{i\phi_{\sigma}}c_{n+1,\sigma}^{\dagger}c_{n,\sigma}^{\dagger}) + h.c.]  \nonumber \\&
-\displaystyle\sum_{1\le n\le L\atop \sigma=1,2} \mu c_{n,\sigma}^{\dagger}c_{n,\sigma}
-t^{\prime}\displaystyle\sum_{1\le n\le L}[c_{n,1}^{\dagger}c_{n,2}+h.c.],
\end{align}
where $c^{\dagger}_{n,\sigma}$($c_{n,\sigma}$) are creation (annihilation) operators on
the ladder ($1\le n\le L$) with $\sigma=1,2$ labeling  
the two legs.  The hopping amplitude in each Kitaev chain is $t$. The inter-leg 
hopping in the superconducting ladder~(S) is $t^{\prime}$ and the nearest-neighbor pairing with phase factor included
is $\Delta e^{i\phi_{\sigma}}$ (in leg $\sigma$ of the ladder) with $\phi_{\sigma}=(-1)^{\sigma}\phi/2$. 
The full Hamiltonian is given by 
\begin{align}
\label{eq:fullham}
H=H_{N_1} + H_{N_1 S} + H_S + H_{N_2 S} + H_{N_2},
\end{align}
where
\begin{align}
H_{N_1 S} = -t'' [c_{0}^{\dagger}c_{1,1}+h.c.]
\end{align}
and 
\begin{align}
H_{N_2 S} = -t'' [c_{L+1}^{\dagger}c_{L,1}+h.c.]
\end{align}
are the terms that couple the superconducting ladder to the metallic leads at the two ends with hopping strength $t''$. 
\begin{figure}
\includegraphics[scale=0.336]{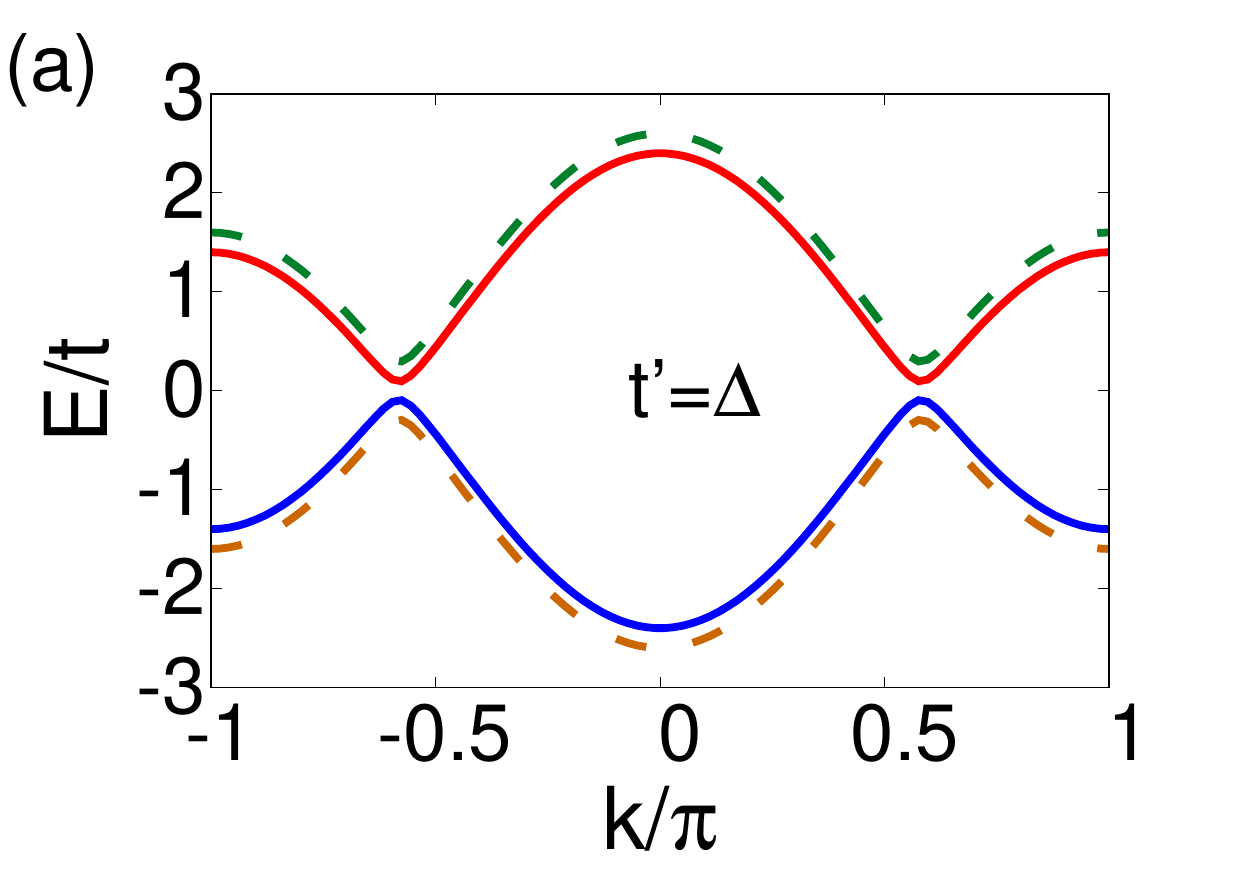}
\includegraphics[scale=0.336]{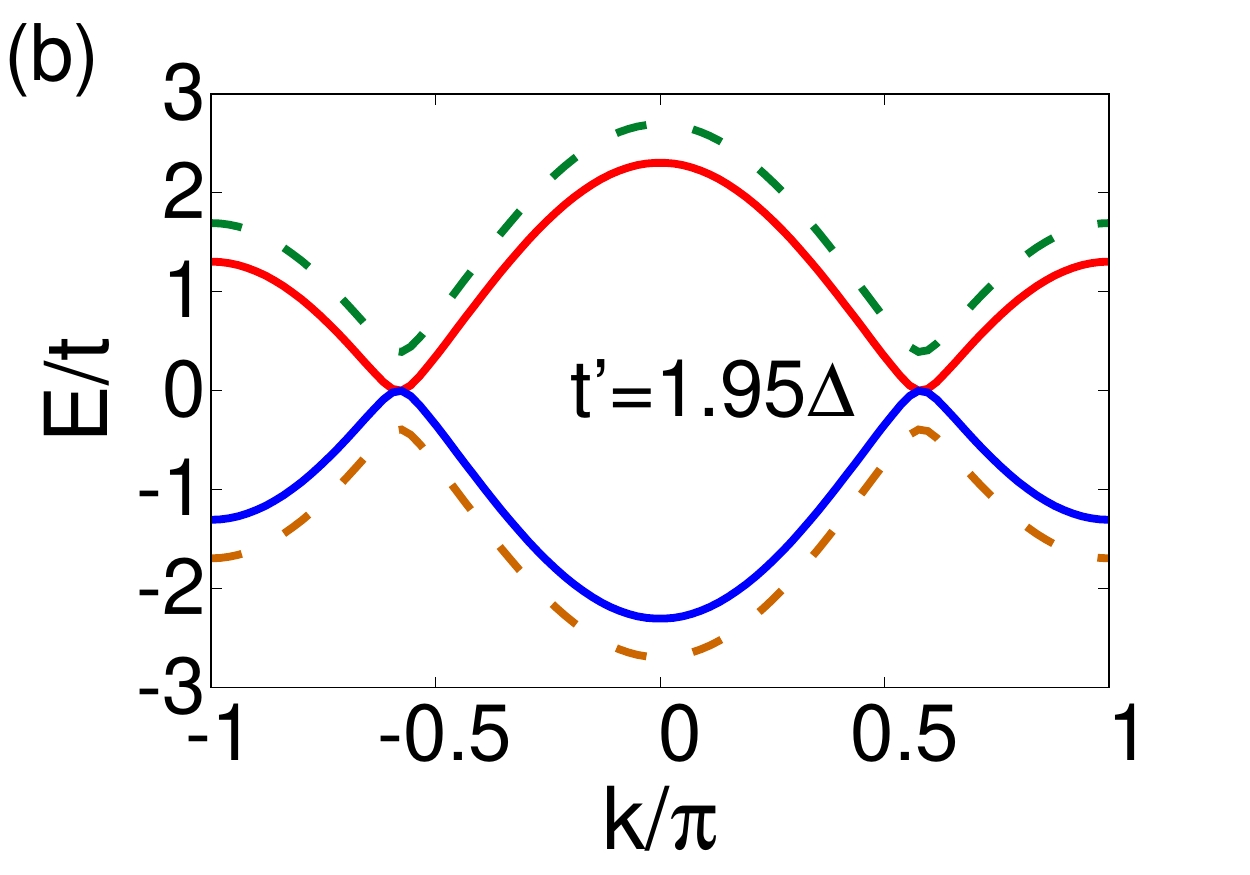}\\
\includegraphics[scale=0.336]{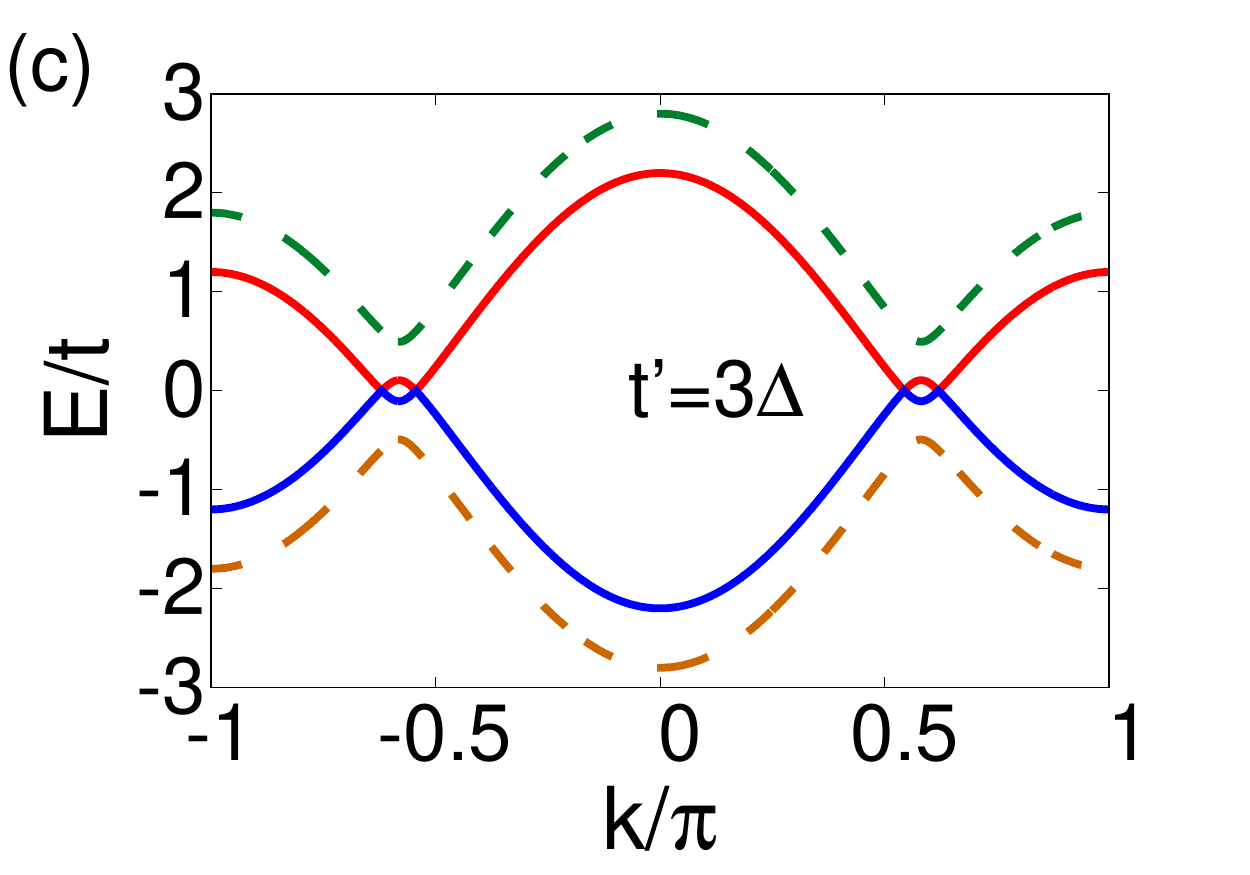}
\caption{The dispersion  of Kitaev ladder with $\phi=\pi$, $\Delta=0.1t$, $\mu=0.5t$. The inter-legs hoppings are 
(a)$t^{\prime}=\Delta$, (b)$t^{\prime}=1.95\Delta$, and (c)$t^{\prime}=3\Delta$. 
Two bands cross at $t^{\prime}=1.95\Delta$ for this value $\mu$ (Eq.~\ref{eq:gap}). The solid curves correspond to $\nu_2=-1$ in Eq.~\eqref{eq:disp-ladder}
while the dashed curves correspond to  $\nu_2=+1$.}
\label{fig:2}
\end{figure}
The dispersion in the ladder region is 
\begin{equation}
\label{eq:disp-ladder} 
E=\nu_1\sqrt{\epsilon_k^2+{t^{\prime}}^2+\alpha_k^2+\nu_2 \cdot2t^{\prime}\sqrt{\epsilon_k^2+\alpha_k^2~\sin^2 \frac{\phi}{2}}},
\end{equation}
where $\nu_1,\nu_2=\pm 1$ correspond to bands formed due to the hybridization of electron and hole excitations in the 
two legs of the ladder, $\phi=(\phi_2-\phi_1)$, $\epsilon_k=-(2t\cos{ka}+\mu)$ and
$\alpha_k=2\Delta\sin{ka}$. The dispersion here looks almost identical
to the one for the $s$-wave superconducting
ladder~\cite{soori2017enhancement}, except for the appearance of the
$k$-dependent part within $\alpha_k$. This spectrum yields four energy
bands as can be seen in Fig.~\ref{fig:2} for $\mu=0.5t$.  The
multiplicity of two for these bands corresponds to bonding and
anti-bonding states formed by the hybridization of the two legs of the
Kitaev ladder while another factor of two corresponds to Bogoliubov-de
Genne~(BdG) quasi-particles formed by the hybridization of electron
and hole bands.

The energy spectrum shows that the gap closes for
$t^{\prime}\ge1.95\Delta$ for $\phi=\pi$ and $\mu = 0.5t$. For a ladder with
$\phi=\pi$, there exist plane wave BdG states at all energies within
the superconducting gap when $t^{\prime}\ge1.95\Delta$. Since the
overall gap in the Kitaev ladder has $k$-dependence
(Eq.~\ref{eq:disp-ladder}), varying $\mu$ shifts the gap-closing
point, in contrast to the s-wave model~\cite{soori2017enhancement}. This motivates the computation
of the energy gap of the ladder as a function of other parameters of the
ladder Hamiltonian. Analytically, it can be shown from
Eq.~\eqref{eq:disp-ladder} that the gap closes only when $\phi=\pi$
and for:
\begin{equation}
 \label{eq:gap}
 t^{\prime} \ge \Delta \sqrt{4-\frac{\mu^2}{t^2-\Delta^2}}~.
\end{equation}
The strongest lower bound here is seen to be $2\Delta$ corresponding to $\mu = 0$.

\begin{figure}
\includegraphics[scale=0.26]{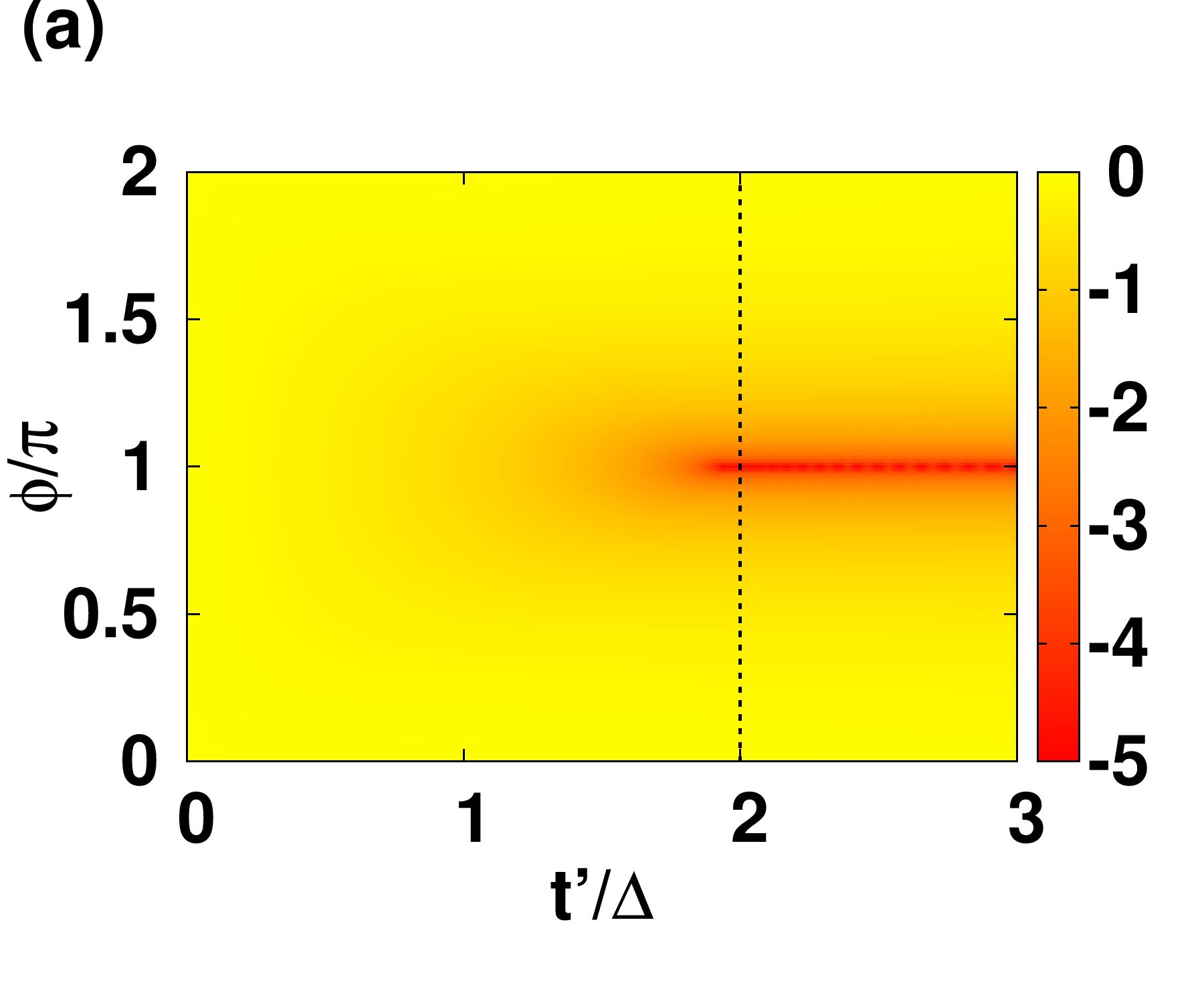}
\includegraphics[scale=0.29]{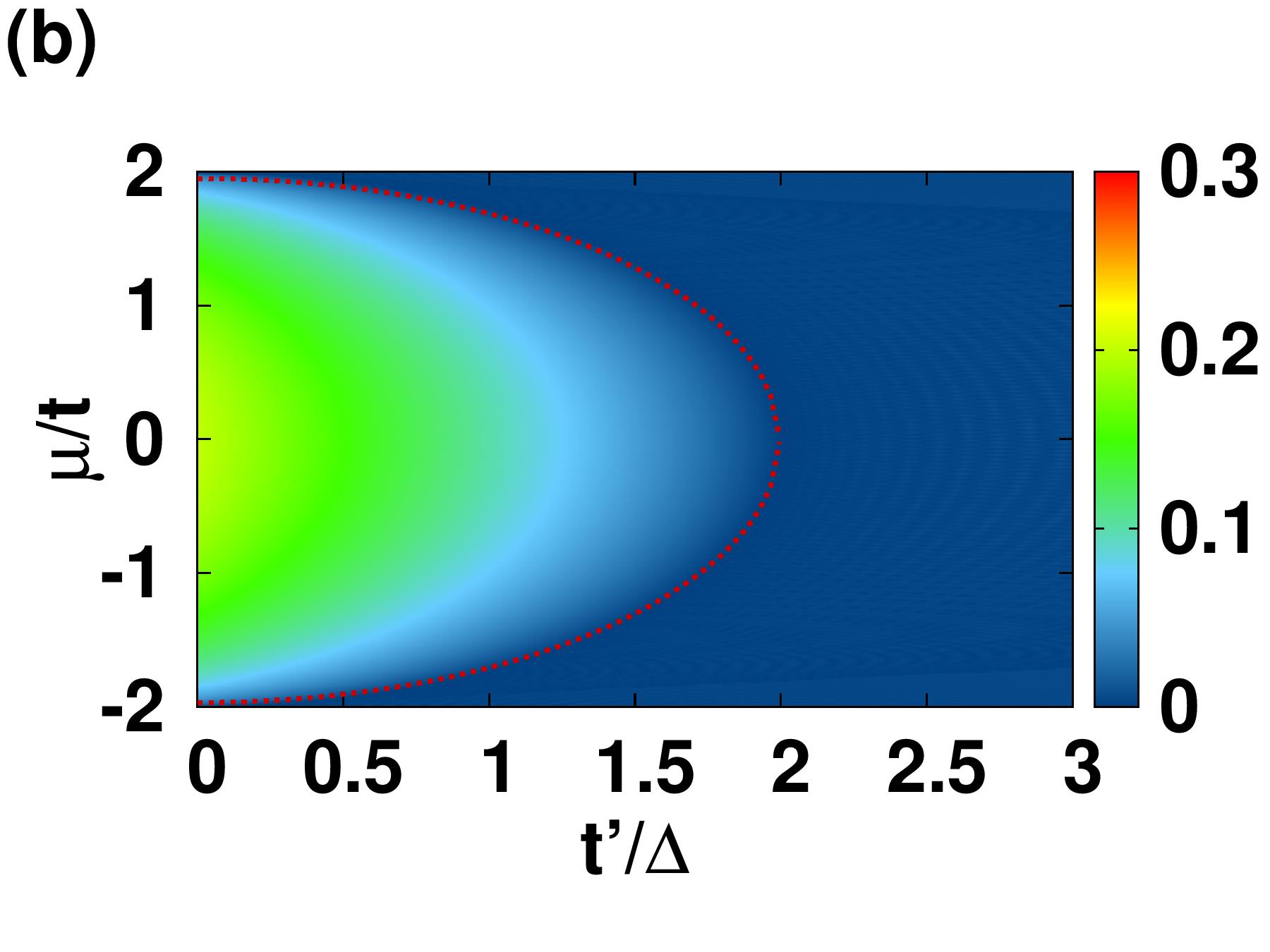}
\caption{ (a)~Logarithm of `the energy gap divided by $4\Delta$' for the ladder with parameters 
$\Delta/t=0.1$,  $\mu/t=0.5$. The dotted line corresponds to $t^{\prime}/\Delta=2$, which is the strongest bound that can be realized when $\mu = 0$ (Eq.~\eqref{eq:gap}).
(b)~The energy gap for the parameters $\Delta/t=0.1$, $\phi=\pi$. The dark line separates gapless 
and gapped regions.}
\label{fig:3}
\end{figure}

When the two Kitaev chains are uncoupled, the spectrum has a gap,
whose maximum value is $4\Delta$ (when the minimum of upper band is at $+2\Delta$
the maximum of the  lower band is at $-2\Delta$).  In the presence of a non-zero
phase difference between the legs, as soon as the inter-leg coupling
$t^{\prime}$ is turned on, plane wave states begin to appear inside this
gap. Such states have a crucial role to play in the transport
properties of the system, and are called subgap Andreev states. For
small inter-leg couplings, despite the presence of subgap Andreev
states, the ladder system still has a gap, although much lower. In
order to quantify this, it is useful to study the logarithm of `the
gap divided by $4\Delta$' as shown in Fig.~\ref{fig:3}~(a) for
$\mu=0.5t$ and $\Delta=0.1t$.  It can be seen that the gap closes for
$\phi=\pi$ and $t^{\prime}\ge 1.95\Delta$ - the closure of the gap
provides further enhancement of transport, as will be described later.
In Fig.~\ref{fig:3}~(b), the gap is plotted as a function of $\mu$ and
$t^{\prime}$ for $\phi=\pi$ and $\Delta=0.1t$. The dark line in the
plot indicates the value of $t^{\prime}$ above which the gap
closes. It can be seen that the gap closes above a critical value of
$t^{\prime}$ which depends on $\mu$ (Eq.~\eqref{eq:gap}).

\section{Wave-functions and Transconductance }\label{section3}
The wavefunction in the metallic regions has the form $[\psi_e,\psi_h]^T$ and in the ladder region it has 
the form  $\Psi=[\psi,\chi]^T$ where $\psi$ and $\chi$ both are two-spinors corresponding to the 
upper and the lower legs of the ladder respectively. For an electron incident from  $N_1$ on 
to the ladder with an energy $E$, the wavefunction takes the following form in the metallic leads:
\begin{equation}
\label{eq:9}\psi_{n,e}=\begin{cases}
\;e^{ik_ean}+r_e e^{-ik_ean}&{\rm for ~~ }n\leq0,\\
\;t_e e^{ik_ean}&{\rm for ~~}n\geq L+1,
\end{cases}
\end{equation}
\begin{equation}
\label{eq:10}\psi_{n,h}=\begin{cases}
\;r_he^{ik_han}&{\rm for~~}n\leq0,\\
\;t_h e^{-ik_han}&{\rm for ~~}n\geq L+1,
\end{cases}
\end{equation}
where $k_{e/h}a=\cos^{-1}\big(\frac{E\pm\mu}{2t}\big)$ and $r_e$, $r_h$, $t_e$, $t_h$ are 
the amplitudes for electron reflection, Andreev reflection, electron tunneling and cross Andreev
reflection respectively. Here, $a$ is the lattice constant and $\hbar k_{e/h}$ is the electron/hole
momentum. The wavefunction in the ladder region takes the form:
\begin{equation}
\label{eq:11}\psi_n=\displaystyle\sum_{\lambda,\nu,p}C_{\lambda,\nu,p}e^{i\lambda k_{\nu,p}an}
[\psi_{e,\lambda,\nu,p}~,~\psi_{h,\lambda,\nu,p}],
\end{equation}
\begin{equation}
\label{eq:12}\chi_n=\displaystyle\sum_{\lambda,\nu,p}C_{\lambda,\nu,p}e^{i\lambda k_{\nu,p}an}
[\chi_{e,\lambda,\nu,p},\chi_{h,\lambda,\nu,p}]
\end{equation}
for $1\leq n\leq L$, where $\lambda=\pm 1$ refers to forward/backward motion of BdG 
quasiparticles, $\nu=\pm 1$ refer to anti-bonding/bonding bands and $p=\pm1$ refers to
electronlike/holelike bands. At a given energy $E$, $k_{\nu,p}a$ is found by numerically solving the 
quartic equation for $\cos{k_{\nu,p}a}$ which is obtained by manipulating Eq.~\eqref{eq:disp-ladder}, 
and the spinor $[\psi_{e,\lambda,\nu,p},~\psi_{h,\lambda,\nu,p},~\chi_{e,\lambda,\nu,p},~\chi_{h,\lambda,\nu,p}]^T$
is the eigenspinor of the ladder Hamiltonian in momentum space with energy $E$ and momentum $\lambda\hbar k_{\nu,p}$.
% \begin{equation}
% \label{eq:13}\cos k_{\nu,p}=\frac{-\mu t+p\sqrt{(t^2-\Delta^2)((E+\nu t^{\prime})^2-4\Delta^2)+\mu^2\Delta^2}}{2(t^2-\Delta^2)}
% \end{equation}
Here, the normal metal lead is connected by a hopping to the upper leg of the ladder (Fig~\ref{fig:1}). 
From the Hamiltonian (Eq.~\eqref{eq:fullham}) the equation of motion at each site on either side of the
junction can be written down. There are six sites and two equations at each site due to particle-hole nature 
of the equations making it twelve equations totally which are just enough to solve for twelve scattering amplitudes 
in Eqs.~\eqref{eq:9},~\eqref{eq:10},~\eqref{eq:11}, and ~\eqref{eq:12}. The details of this calculation are shown in the Appendix.
% \bea
% E\psi_{e,0}&=&-t\psi_{e,-1}-t\psi_{e,1}-\mu\psi_{e,0} \nonumber \\
% E\psi_{h,0}&=&+t\psi_{h,-1}+t\psi_{h,1}+\mu\psi_{h,0} \nonumber \\
% to be complete
% \eea

A useful quantity to study the relative contribution of CAR with
respect to ET is the differential
transconductance. Also, this is the physical quantity that is measured in 
transport experiments. The ladder attached to two normal metals is biased so that a
voltage $V$ is applied to $N_1$ keeping the ladder and $N_2$ grounded.
The differential transconductance, $G_{21}:=\frac{dI_2}{dV_1}$ is the
ratio of change in the current $dI_2$ in $N_2$ to the change in the
applied voltage $dV_1$ in $N_1$. From Landauer-B\"uttiker
formalism~\cite{landauer1957r,landauer1970r,buttiker1985m,buttiker1986m,datta1995electronic},
the differential transconductance of the system at bias $V_1$ is given by
\begin{equation}\label{eq:13}
G_{21}=\frac{e^2}{h}\Bigg(|t_e|^2-|t_h|^2\frac{\sin{k_h a}}{\sin{k_e a}}\Bigg).
\end{equation}
Here, the first term represents the contribution to transconductance from ET while the second term represents 
the contribution due to CAR. Therefore, a positive $G_{21}$ is a clear signature of enhanced ET while 
a negative $G_{21}$ is a signature of enhanced CAR.

% The other quantity of interest is charge on sites of the Kitaev ladder(S). The charge on each site of ladder is simply the difference of probabilities of quasiparticles on that site. The charge on $n^{th}$($1\le n\le L$) site of ladder(S) is
% \begin{equation}
% \label{eq:15}
% Q_n=e(|\psi_{e,n}|^2-|\psi_{h,n}|^2)
% \end{equation}
% where,$e$ is the unit charge of an electron. This implies negative charge(in units of $e$) on a site signifies the presence of hole and positive(in units of $e$) charge represents the electron dominant on that site.
\section{Results and analysis}
\subsection{Variation of $\mu$}

One special case where crossed Andreev reflection happens is when $\phi = 0$.  In this case,
the two legs of the ladder retain their Majorana fermions and crossed
Andreev reflection can happen by the non-local state formed by the
coupling between two Majorana bound states at the end. But this
carries no net current due to the competing electron tunneling which
has a magnitude of current that is same as that of crossed Andreev
reflection~\cite{nilsson2008splitting}. In general, we find that choosing $\phi = \pi$ and $t''=t$ works best for the enhancement of
CAR and ET, and therefore fix $\phi = \pi$ and $t''=t$ in this paper, unless
specified otherwise. 
\begin{figure}[h!]
\includegraphics[scale=0.28]{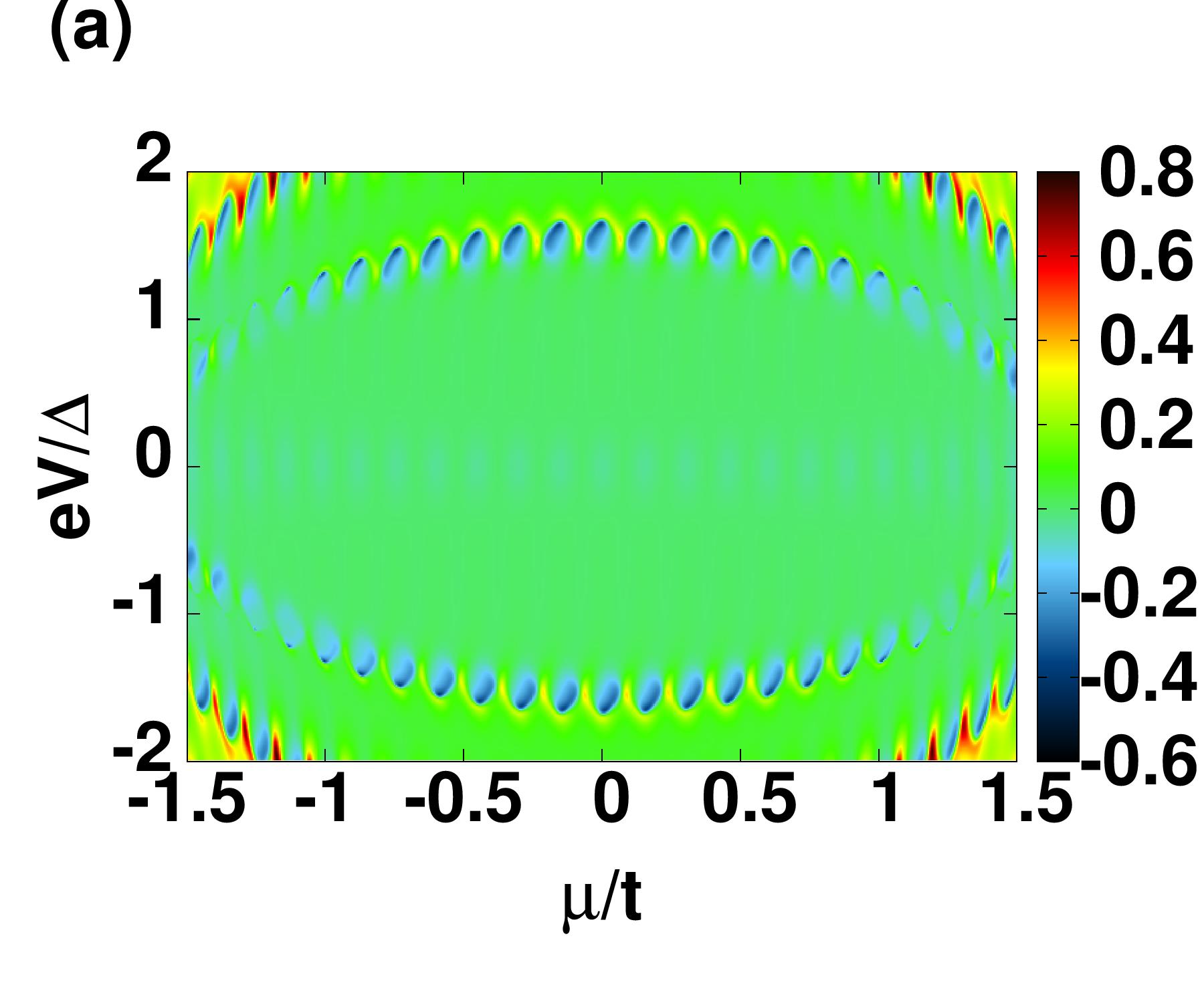}
\includegraphics[scale=0.28]{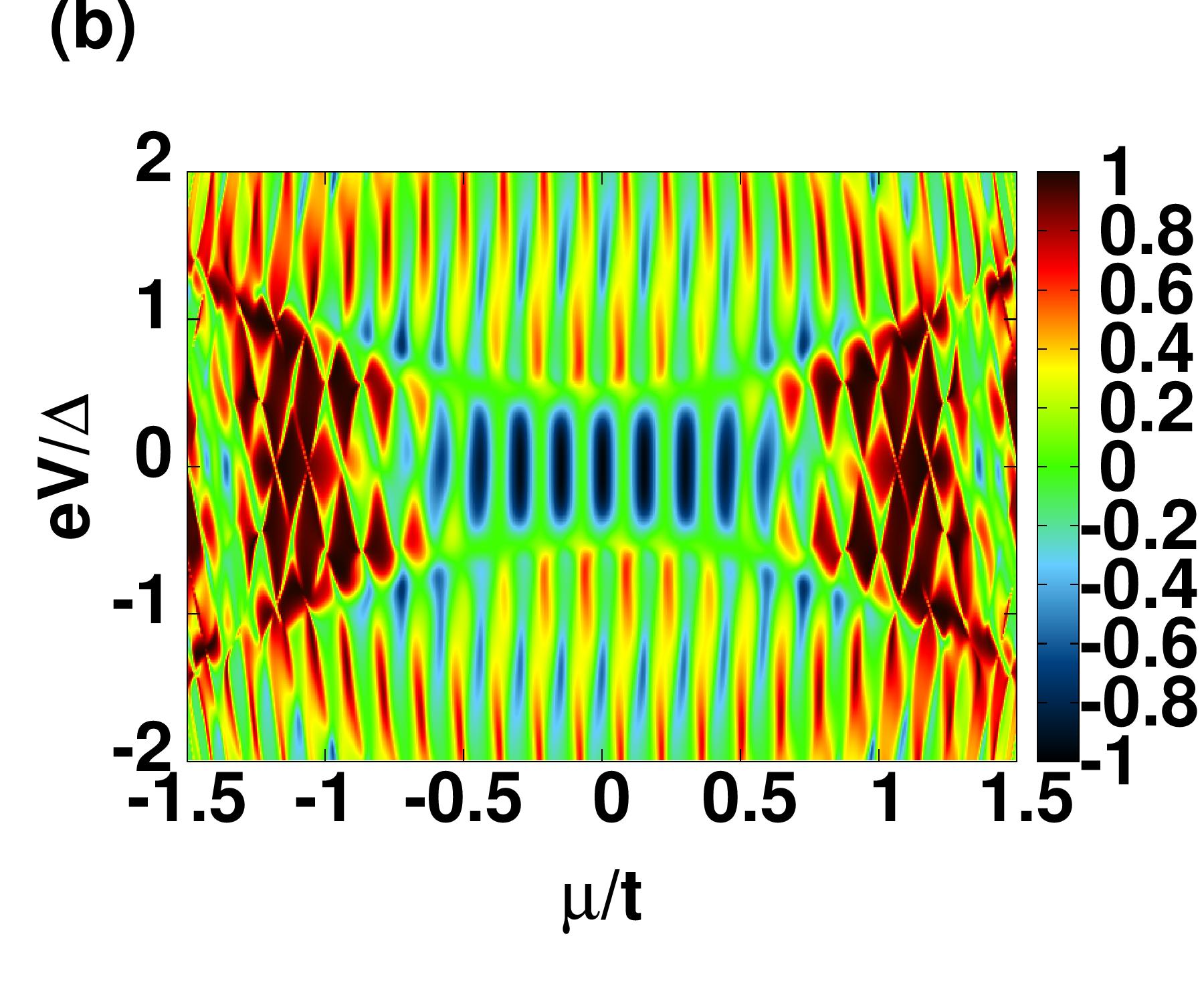}
\caption{The transconductance $G_{21}$ in units of $e^2/h$ for the parameters:
$\Delta=0.1t$, $t''=t$, $\phi=\pi$, $L=40$ and (a) $t^{\prime}=\Delta$ (b) $t^{\prime}=3\Delta$.
(a) Below the critical value of inter-leg hopping, the transconductance is mainly zero 
 while (b) beyond the critical value an enhancement of both CAR and ET can be 
seen due to the presence of subgap Andreev states.   }
\label{fig:4}
\end{figure}

In Fig.~\ref{fig:4} we plot the differential transconductance as a
function of bias and chemical potential for two values of the inter-leg
hoppings: (a)~$t^{\prime}=\Delta$ and (b)~$t^{\prime}=3\Delta$. We see
that the transconductance is mostly suppressed for the case
$t^{\prime}<2\Delta$, while for the case $t^{\prime}>2\Delta$, one can
find thick regions in the plot where the transconductance is
enhanced. Enhancement of differential transconductance in magnitude is
due to the existence of subgap Andreev states.

\begin{figure}[h!]
 \includegraphics[width=8cm]{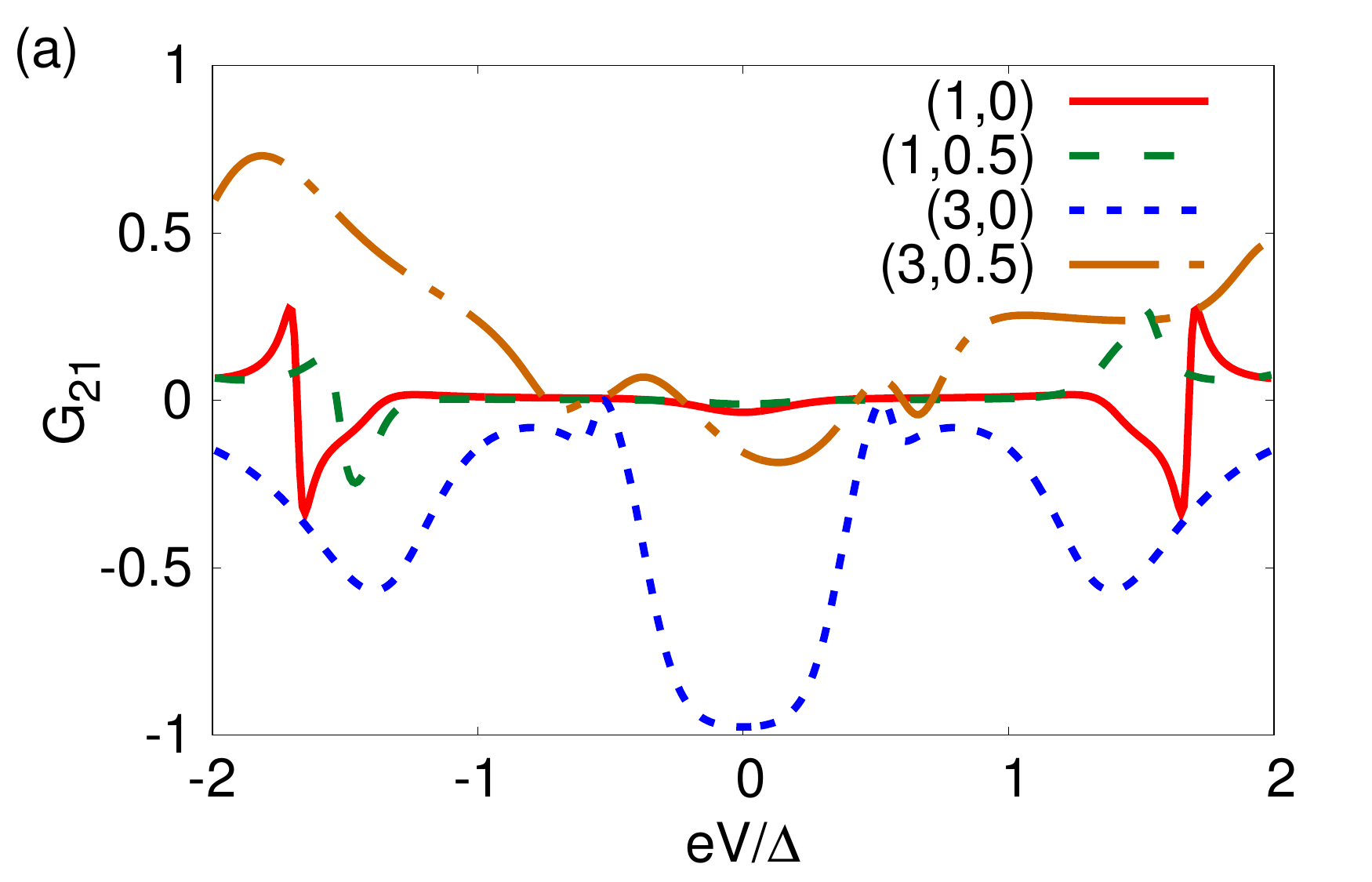}
 \includegraphics[width=8cm]{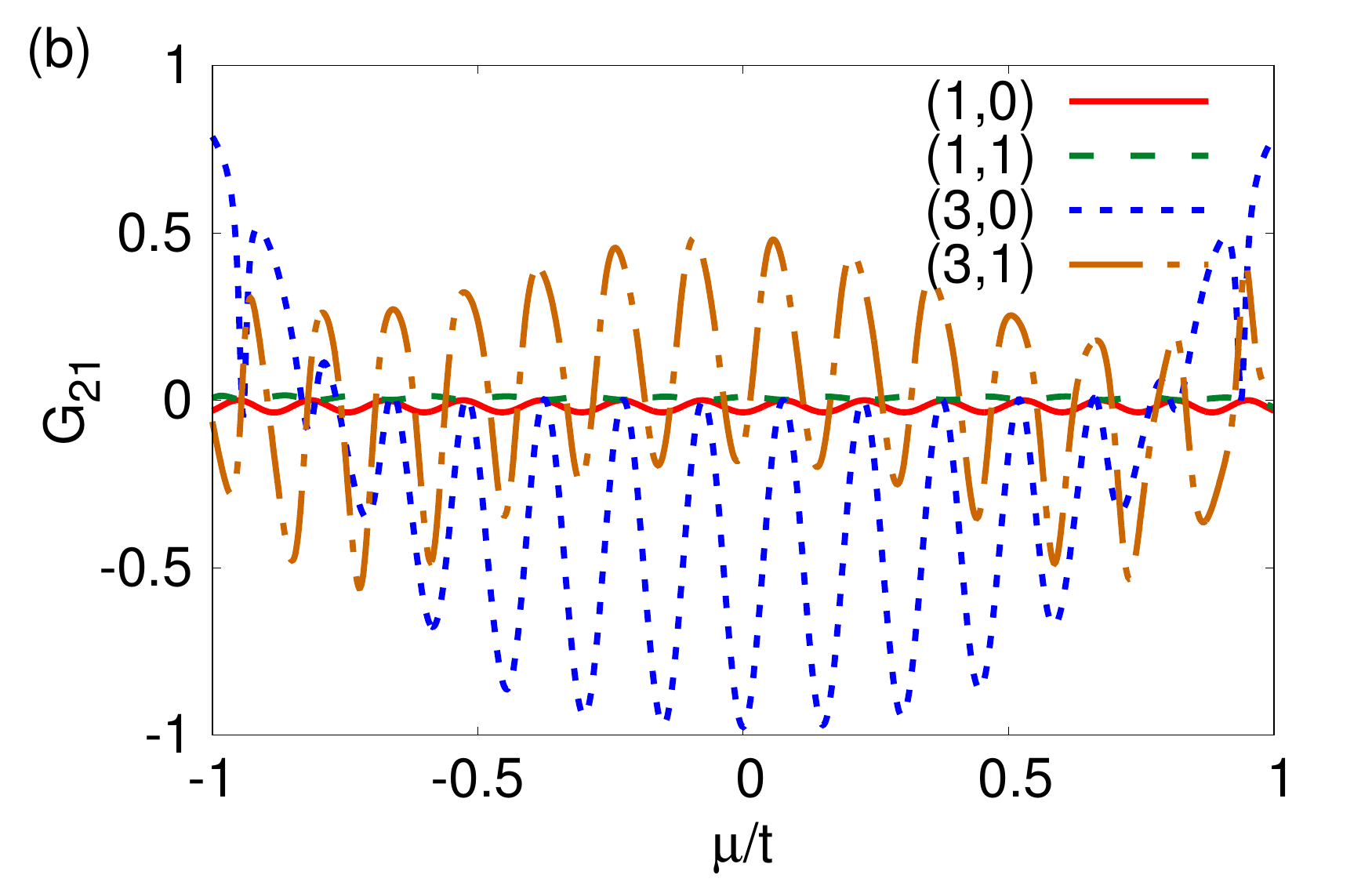}
 \caption{The transconductance $G_{21}$ in units of $e^2/h$ as a function of (a) bias voltage and 
 (b) chemical potential (some of the results in Fig.~\ref{fig:4} plotted as line plot). The legend 
 shows (a)~$(t'/\Delta,\mu/t)$, (b)~$(t'/\Delta,eV/\Delta)$ for different curves.}
 \label{fig:5}
\end{figure}

Some of the results in Fig.~\ref{fig:4} are replotted in Fig.~\ref{fig:5}
for clarity.
It can be seen from both the plots that the magnitude of transconductance is high for $t'=3\Delta$.
In Fig.~\ref{fig:5}(a), it can be seen that $G_{21}$ is not a symmetric function of $eV$ for $\mu=0.5t$, while 
$G_{21}$ is a symmetric function of $eV$ for $\mu=0$. In Fig.~\ref{fig:5}(b), one can see periodic oscillations 
of transconductance as a function of $\mu$, which we will soon touch upon.

Before we analyze the origin of enhanced CAR and ET, let us examine the contribution
of various processes in the transport. An incident electron can do one of four things:
reflect back (ER -electron reflection), reflect back as a hole (AR -Andreev reflection),
transmit through and emerge out as electron (ET -electron tunneling) or transmit through
and emerge out as a hole (CAR -crossed Andreev reflection). Conservation of probability 
currents for these processes implies~\cite{btk,soori2013transport}
\beq \label{eq:prcons}|r_e|^2+|t_e|^2+ (|r_h|^2+|t_h|^2)\frac{\sin{k_h a}}{\sin{k_e a}} = 1. \eeq
We now examine these components for the set of parameters as in Fig.~\ref{fig:4}(a) and~\ref{fig:4}(b).
\begin{figure}[h!]
 \includegraphics[width=8cm]{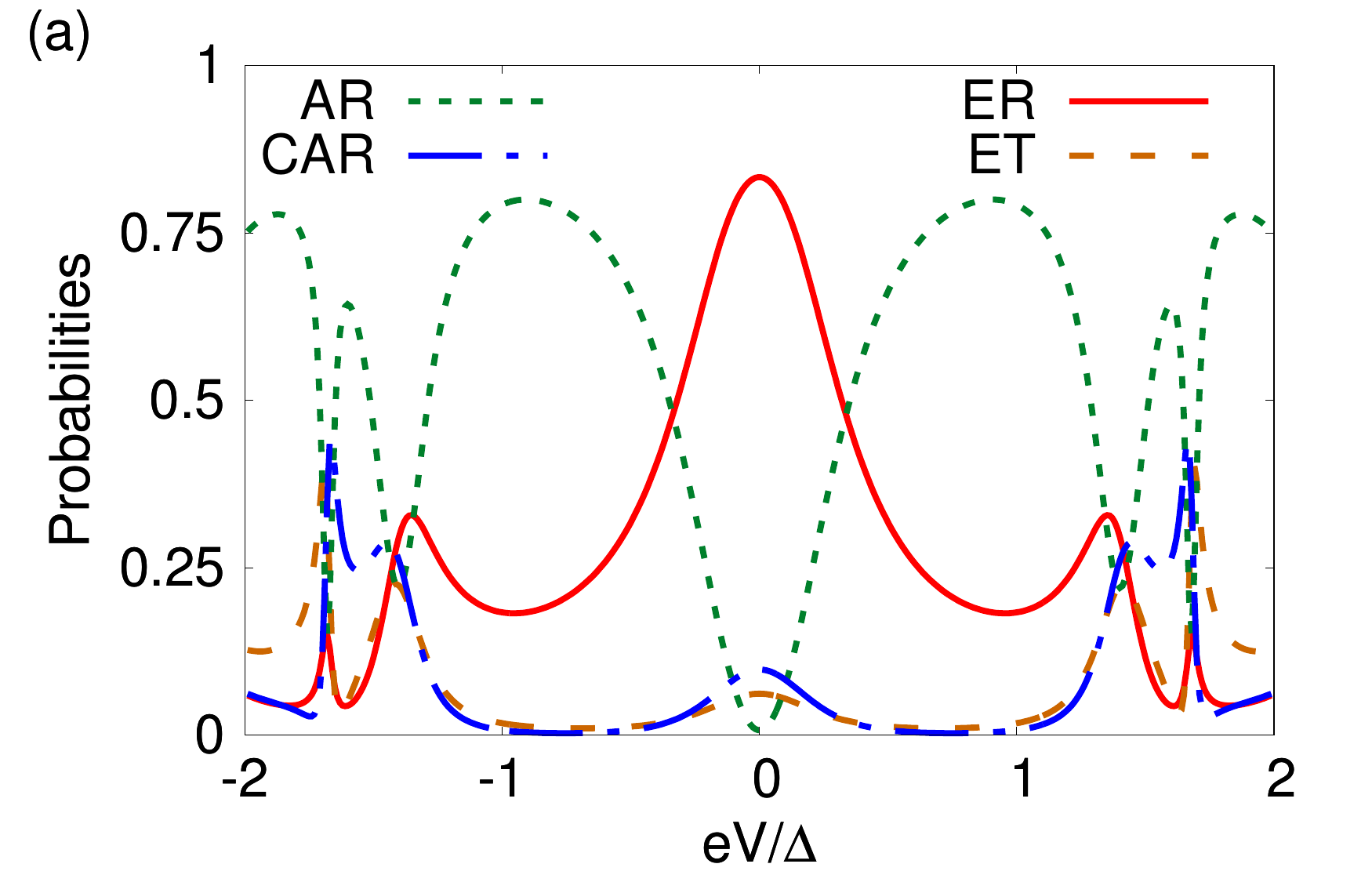}
 \includegraphics[width=8cm]{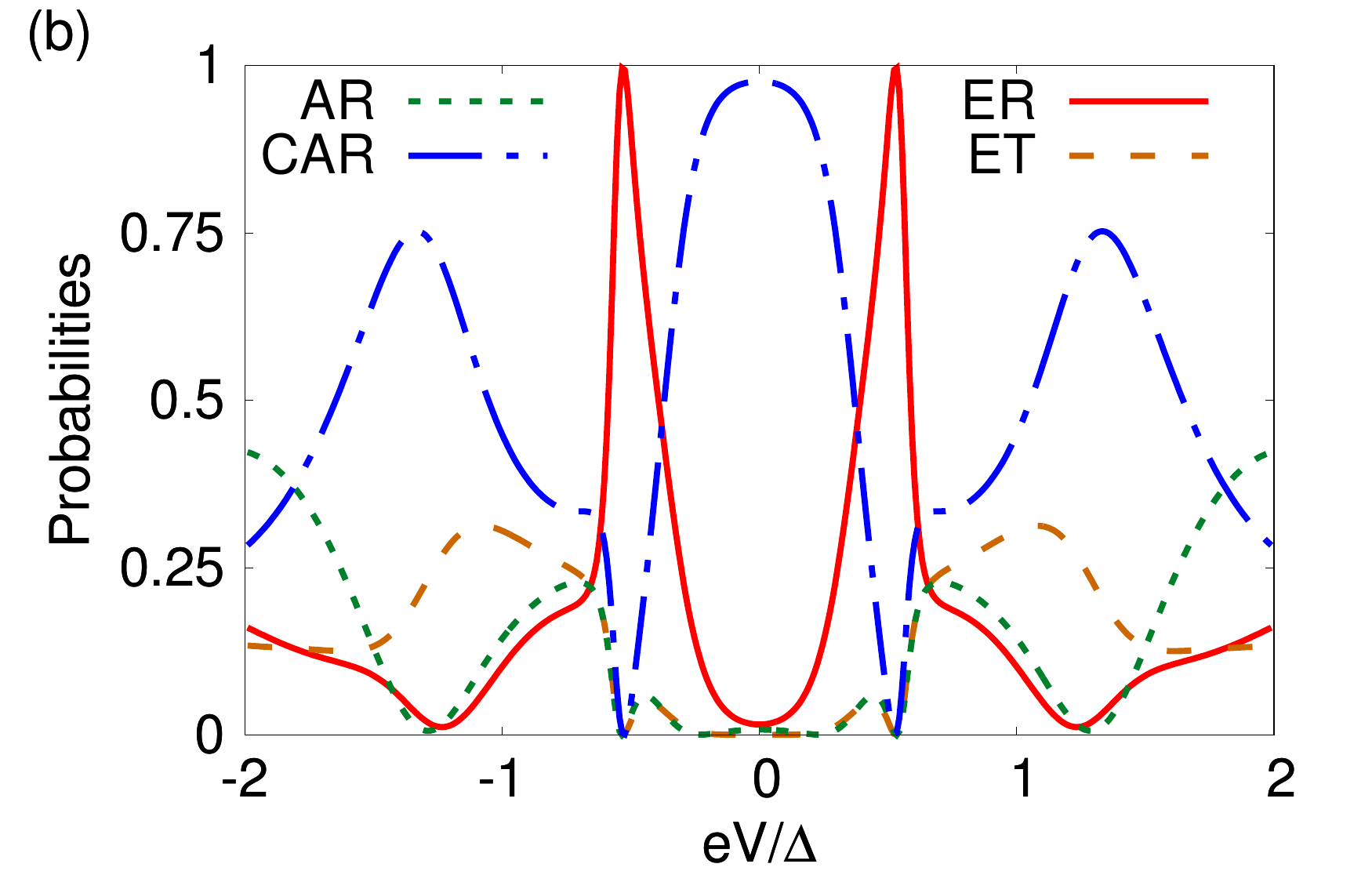}
 \caption{The probability of four processes as shown in Eq.~\eqref{eq:prcons} as a function of applied bias for parameters (a)$t'=\Delta$, (b) $t'=3\Delta$ same as Fig.~\ref{fig:4} and $\mu=0$. It can be seen that (a) below the critical value ER and AR dominates while (b) going beyond the critical value CAR and ER dominates with CAR touching its maximum value around the zero bias voltage.  
 }
 \label{fig:6}
 \end{figure}
 It can be seen from Fig.~\ref{fig:6}(a) that when $t'=\Delta$, ER and AR dominate the transport.
%\begin{figure}[h!]
%\includegraphics[width=8cm]{fig_5bmu0-eps-converted-to.pdf}
% \caption{Four terms in eq.~\eqref{eq:prcons} as a function of applied bias for parameters 
% same as Fig.~\ref{fig:5}(b) and $\mu=0$.
 %}\label{fig:line-5bmu0}\end{figure}
 It can be seen from Fig.~\ref{fig:6}(b) that when $t'=3\Delta$, CAR dominates the transport around
 zero bias. This feature is one of the unique aspects of the present setup. Such CAR-enhanced regions can also be seen in the contour plot of Fig.~\ref{fig:4}(b) where
 $G_{21}$ hits values close to $-e^2/h$.  We now turn to the dependence of various probabilities on $\mu$ in the zero bias case. 
 \begin{figure}[h!]
 \includegraphics[width=8cm]{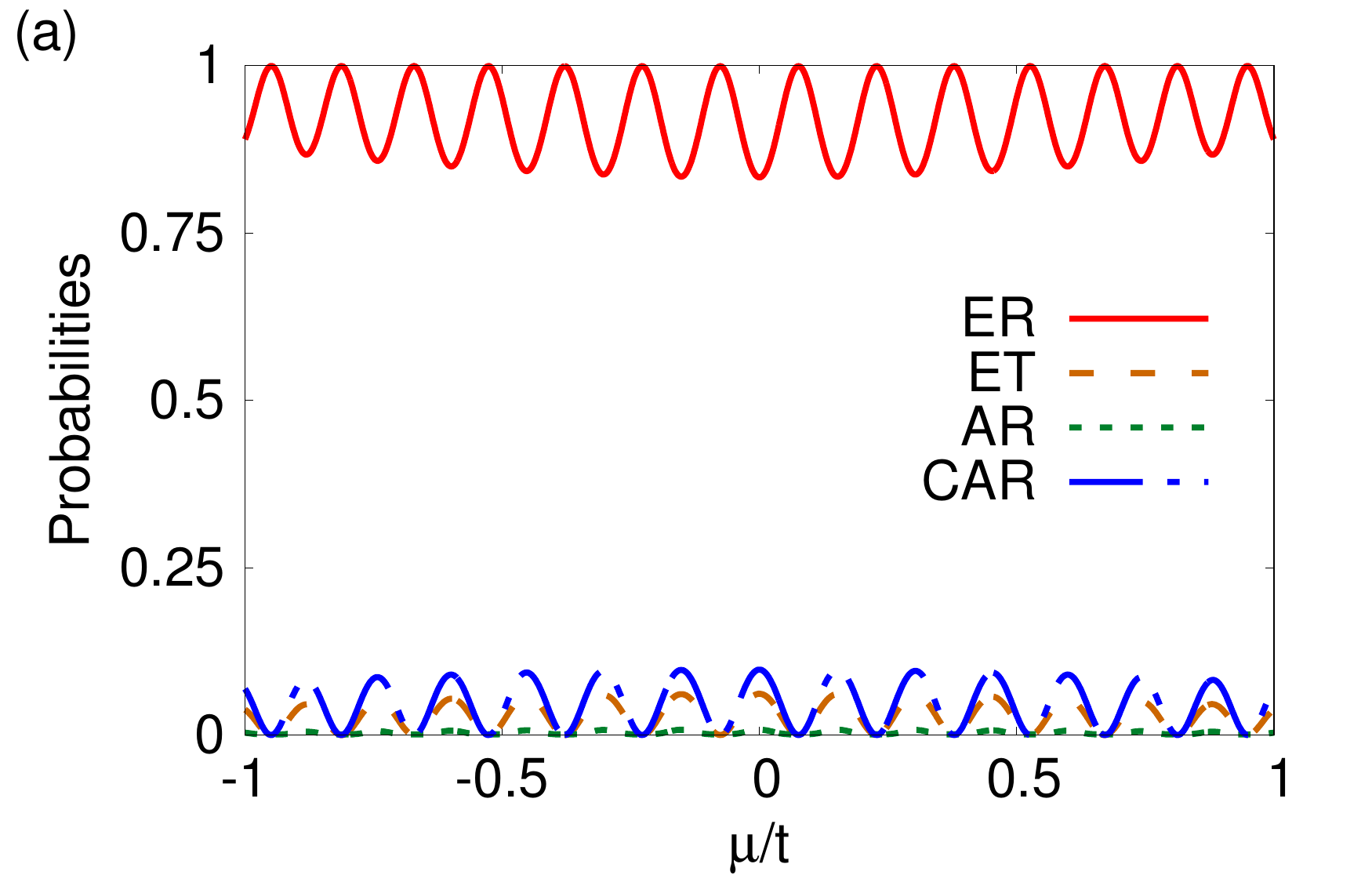}
 \includegraphics[width=8cm]{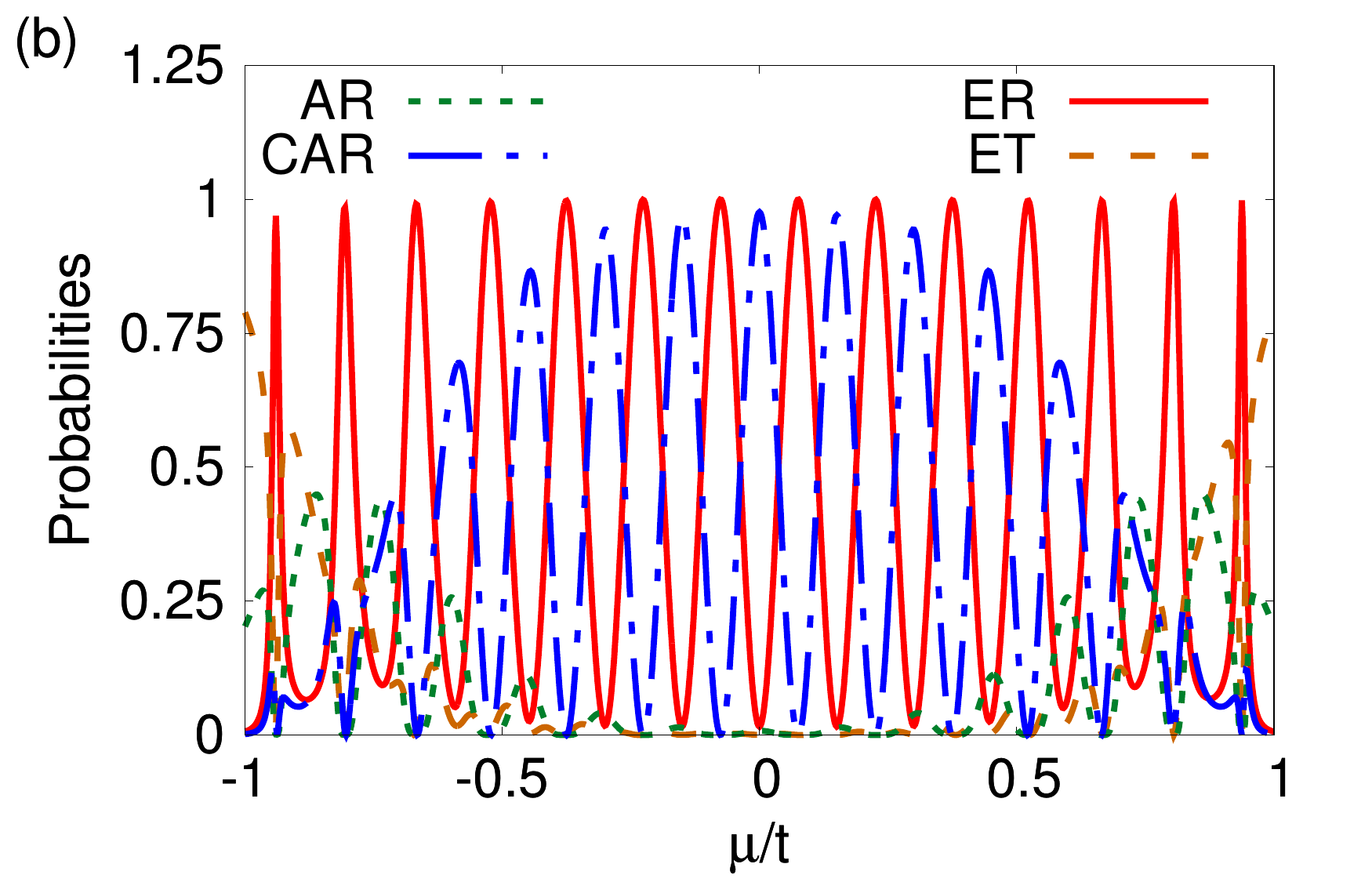}
 \caption{The probability of various processes in system as a function of chemical potential for (a) $t'=\Delta$ and (b) $t'=3\Delta$ with other parameter same as Fig.~\ref{fig:4} and $eV=0$. Here, CAR dominates over other processes after crossing of bands when $t'\geq 2\Delta$.
 }
 \label{fig:7}
 \end{figure}
 From Fig.~\ref{fig:7}(a), we can see that for $t'=\Delta$, ER dominates and the other three process are 
 suppressed at zero bias. 
% \begin{figure}[h!]
% \includegraphics[width=8cm]{fig_5bV0.eps}
% \caption{Four terms in eq.~\eqref{eq:prcons} as a function of chemical potential for parameters 
 %same as Fig.~\ref{fig:5}(b) and $eV=0$.
 %}\label{fig:line-5bV0}\end{figure}
 From Fig.~\ref{fig:7}(b), we can see that for $t'=3\Delta$, ER and CAR are enhanced. Whether ER is enhanced 
 or CAR is enhanced depends on the exact value of $\mu$ and one is enhanced at the expense of the other while suppressing 
 ET and AR. In non-local conductance measurements, only $G_{21}$ which is a linear combination of CAR and ET probabilities can be 
 measured; the two probabilities cannot be separately measured. 

\begin{figure}[h!]
\includegraphics[scale=0.32]{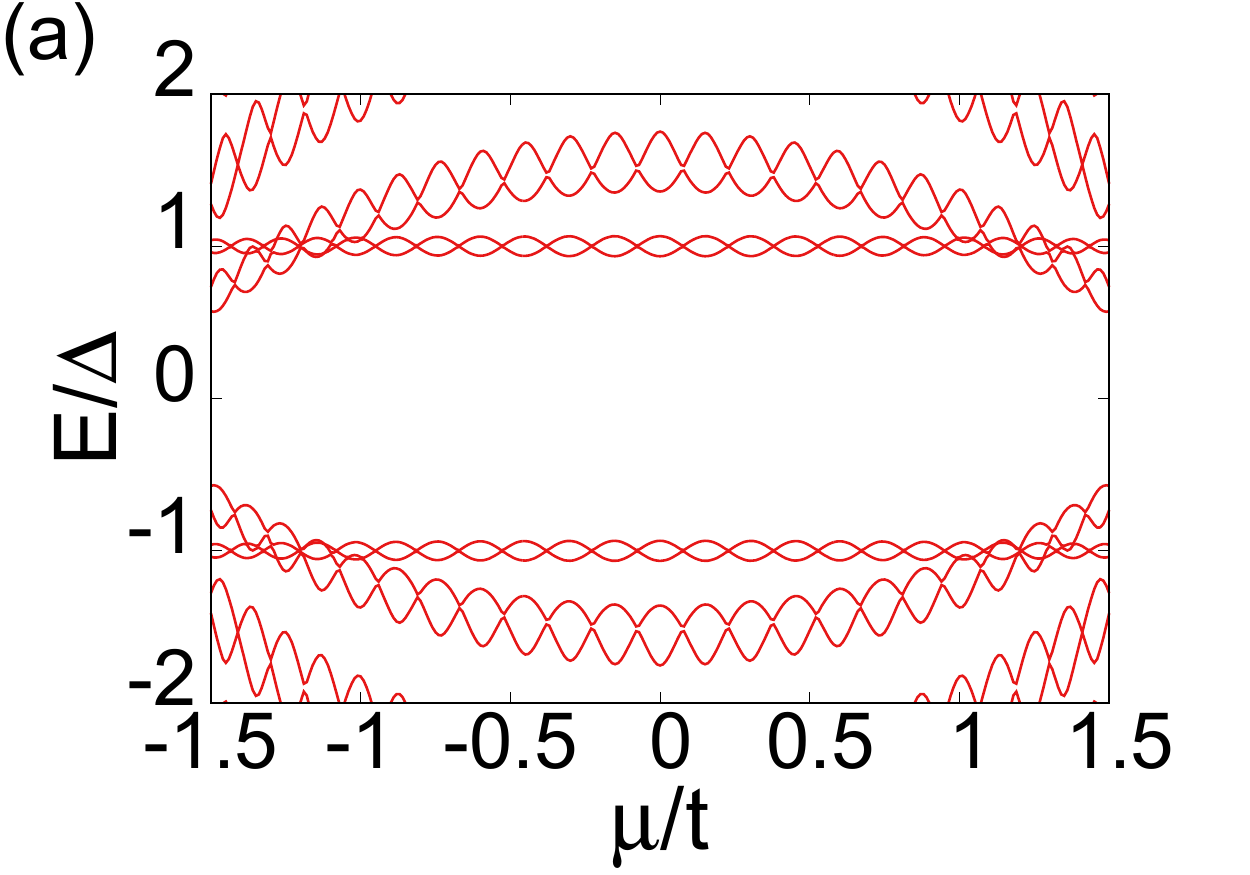}
\includegraphics[scale=0.32]{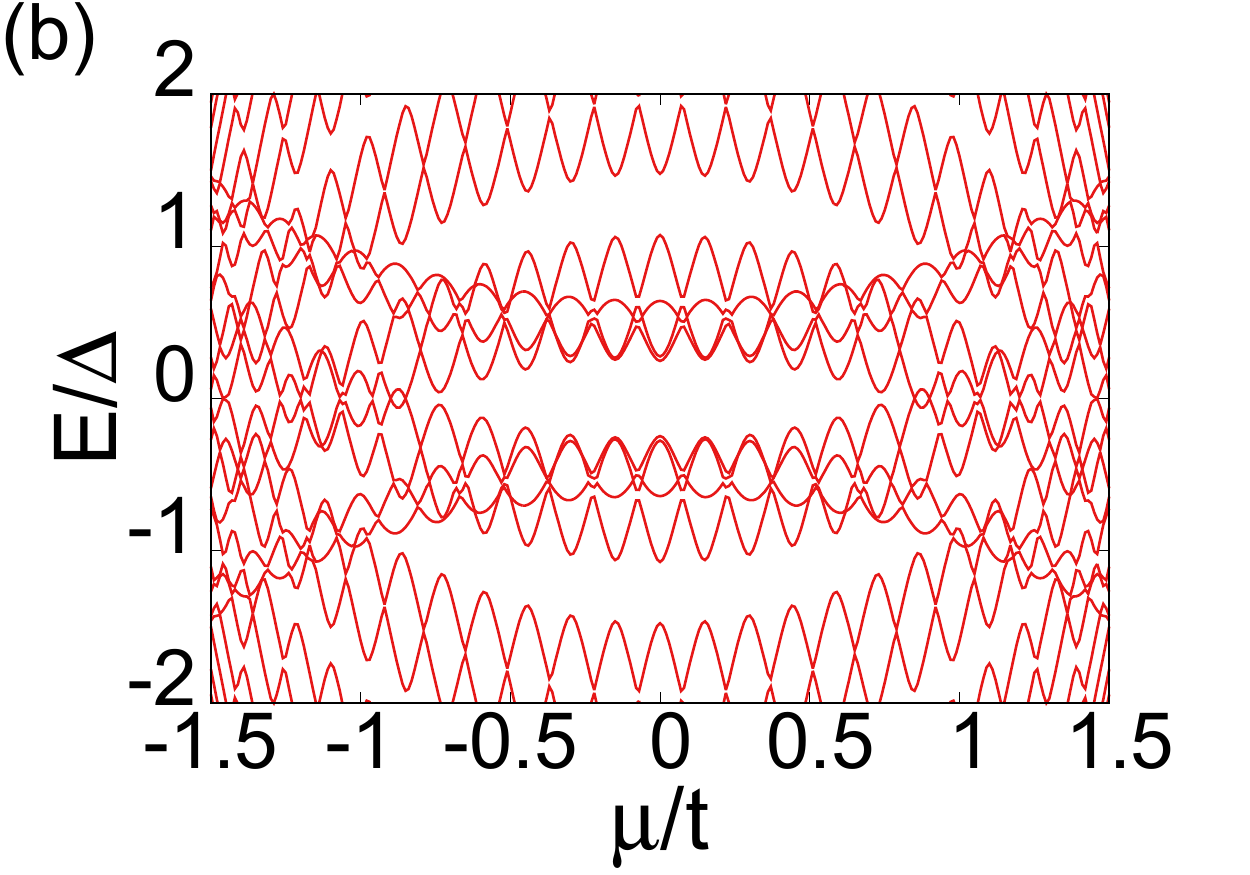}
\caption{The energy spectrum of the isolated Kitaev ladder with open boundary conditions for 
(a) $t^{\prime}=\Delta$, (b) $t^{\prime}=3\Delta$ such that the other parameters are $\Delta=0.1t$,
$\phi=\pi$, $L=40$. The gapped region in (a) shows the sparsely spaced subgap Andreev states
whereas going beyond a critical value of inter-leg hopping these states become dense (b). }
 \label{fig:8}
\end{figure}

In Fig.~\ref{fig:8}, we plot the subgap energy states of the isolated
ladder as a function of $\mu$ for the same parameters as in
Fig.~\ref{fig:4}. The features in Fig.~\ref{fig:4} can be directly
compared with Fig.~\ref{fig:8}.
 We can see a resemblance between the
features of the two plots in Fig.~\ref{fig:4} and the features of the
plots for respective parameters in Fig.~\ref{fig:8}. The resemblance is
a signature of resonant transmission of charge from one reservoir to another
 through a quantum dot where the ladder plays the
role of the quantum dot~\cite{soori2009,soori19}.

In Fig.~\ref{fig:4}(a) (for $t^{\prime}<2\Delta$), the center of the bias window has zero
transconductance, since there are no states available in this region
as confirmed in Fig.~\ref{fig:8}(a). On the other hand, there is high ET and CAR near the
boundary of the window; correspondingly the presence of energy levels
in that region is shown in Fig.~\ref{fig:8}(a).  In Fig.~\ref{fig:4}(b) (for
$t^{\prime}>2\Delta$) the CAR and ET both show enhancement due to the
presence of subgap states as shown in Fig.~\ref{fig:8}(b). These
subgap states provide the plane wave modes which promote the
transmission of quasiparticles. It is seen from Fig.~\ref{fig:4}(b) that both
ET and CAR show periodic behavior with varying chemical
potential ($\mu$). This periodicity in differential transconductance
can be understood as due to Fabry-P\'erot interference of subgap Andreev states in the
ladder region. The Fabry-P\'erot interference condition
$(k_{i+1}-k_i)aL\approx\pi$ gives a spacing of $\delta \mu\approx 0.157t$
between consecutive peaks in transconductance values for the parameters of
Fig.~\ref{fig:4}(b) in the region $eV=0$ and $\mu\approx 0$. This
agrees with the spacing in the transconductance plot of
Fig.~\ref{fig:4}(b).

\begin{figure}[h!]
\includegraphics[scale=0.26]{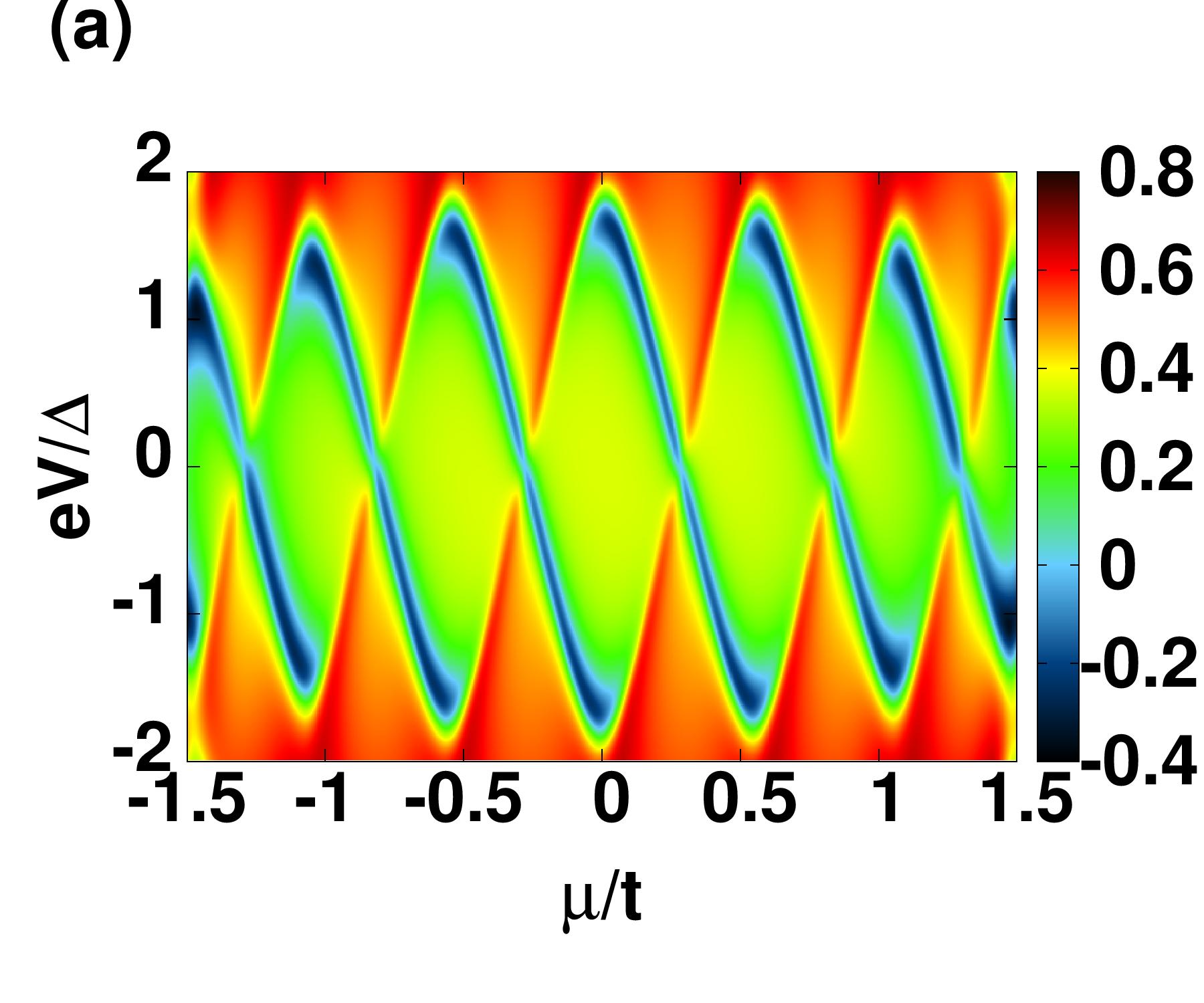}
\includegraphics[scale=0.26]{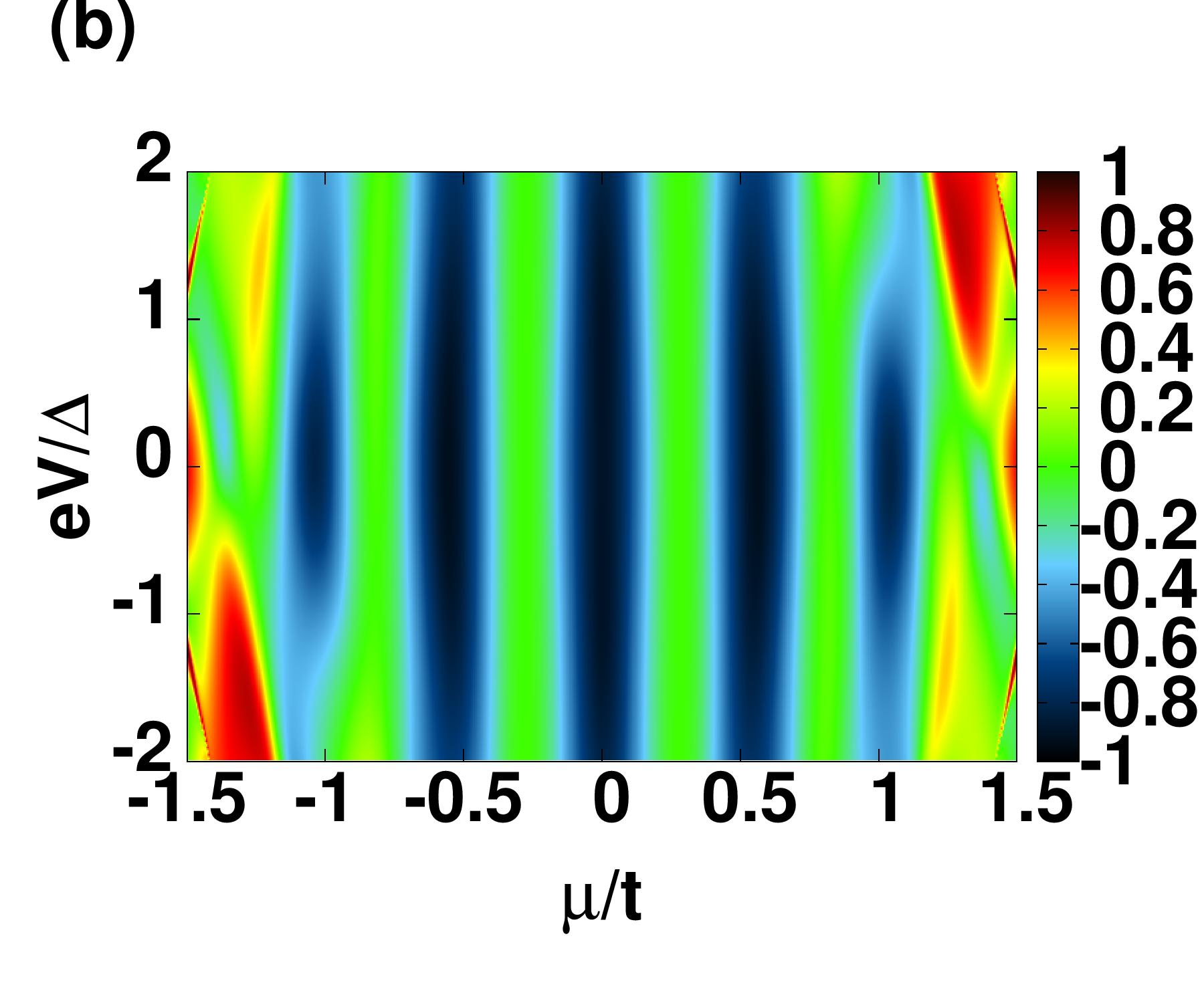}
\caption{The transconductance $G_{21}$ (in units of $e^2/h$) for the
  same parameters as in Fig.~\ref{fig:4} but with a smaller system
  size ($L=10$).  The smaller system size allows for the electrons
  (holes) to tunnel through the junction even though the inter-leg
  hopping is below the critical value (a), whereas for the choice of
  $t^{\prime}$ above the critical value the enhancement of CAR and ET to its
  extreme value is obtained (b).  The parameters are: $\Delta=0.1t$,
  $\mu=0.5t$, $\phi=\pi$ and (a) $t^{\prime}=\Delta$ (b)
  $t^{\prime}=3\Delta$.}
\label{fig:9}
\end{figure}

However, for very small system sizes the transconductance shows
distinctly different behavior as can be seen in Fig.~\ref{fig:9},
which is the same as Fig.~\ref{fig:4} but with $L = 10$. For
$t^{\prime}=\Delta$, the ET is enhanced with a maximum conductance of
$0.8e^2/h$ whereas for $t^{\prime}=3\Delta$ the CAR dominates with an
extremum value of $-e^2/h$  along with enhanced ET mainly at the
corners of the contour plot with a maximum value of $e^2/h$.
\subsection{Variation of $L$}
\begin{figure}[h!]
\includegraphics[scale=0.28]{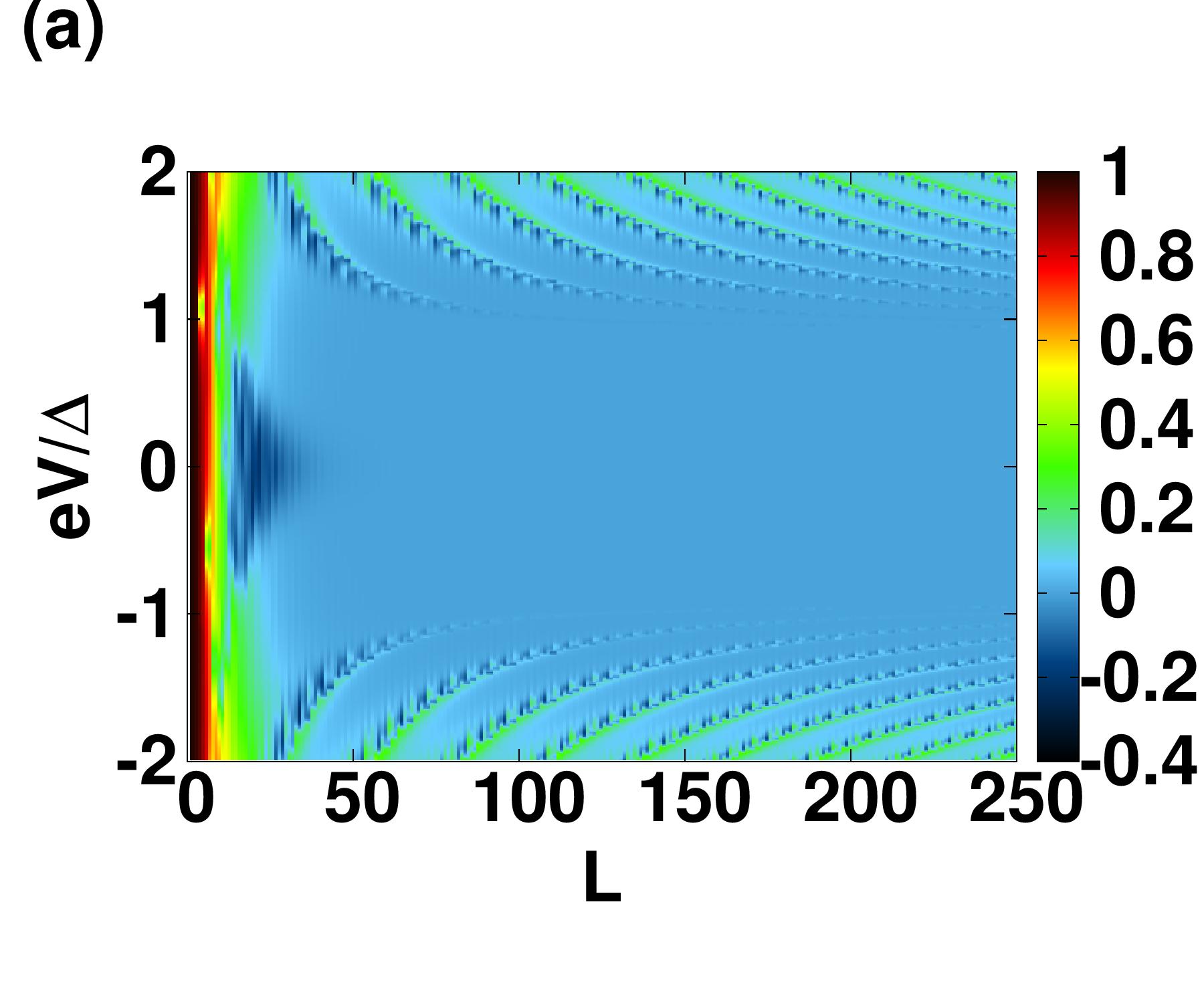}
\includegraphics[scale=0.28]{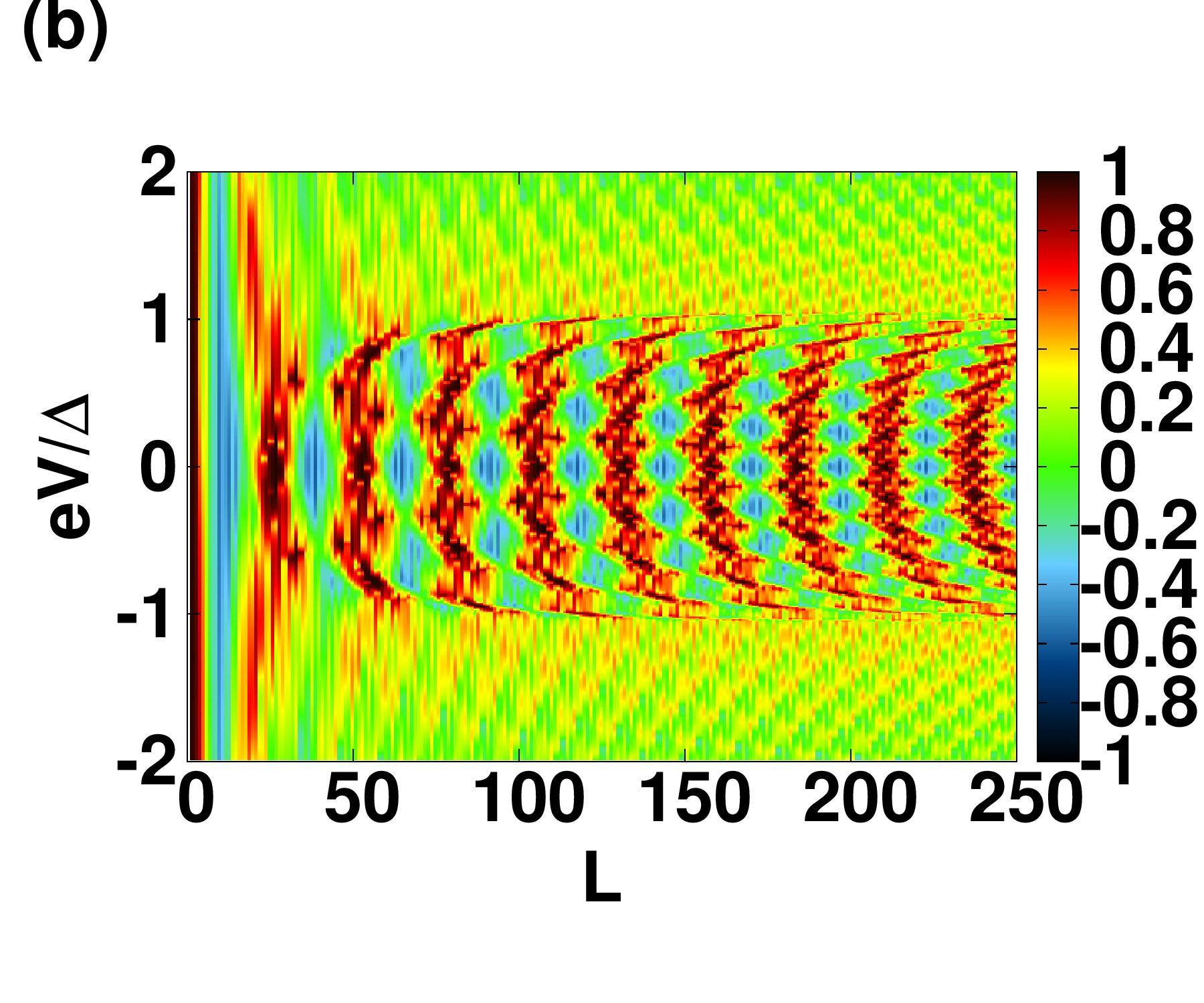}
\caption{The system size variation of transconductance (plotted in units 
of $e^2/h$) in the two regimes: 
(a) below the critical value of inter-leg hopping ($t^{\prime}=\Delta$) and 
(b) above the critical value $t^{\prime}=3\Delta$. The decaying modes in 
(a) suppress transport for larger system sizes for $eV<|\Delta|$, however 
for $eV>|\Delta|$ the plane wave BdG modes are responsible for the transport. 
The periodic nature in (b) appears as a consequence of Fabry-P\'erot resonance. 
The other parameters are: $\Delta=0.1t$,  $\mu=0.5t$, 
$\phi=\pi$.}
\label{fig:10}
\end{figure}

\begin{figure}[h!]
\includegraphics[width=8cm]{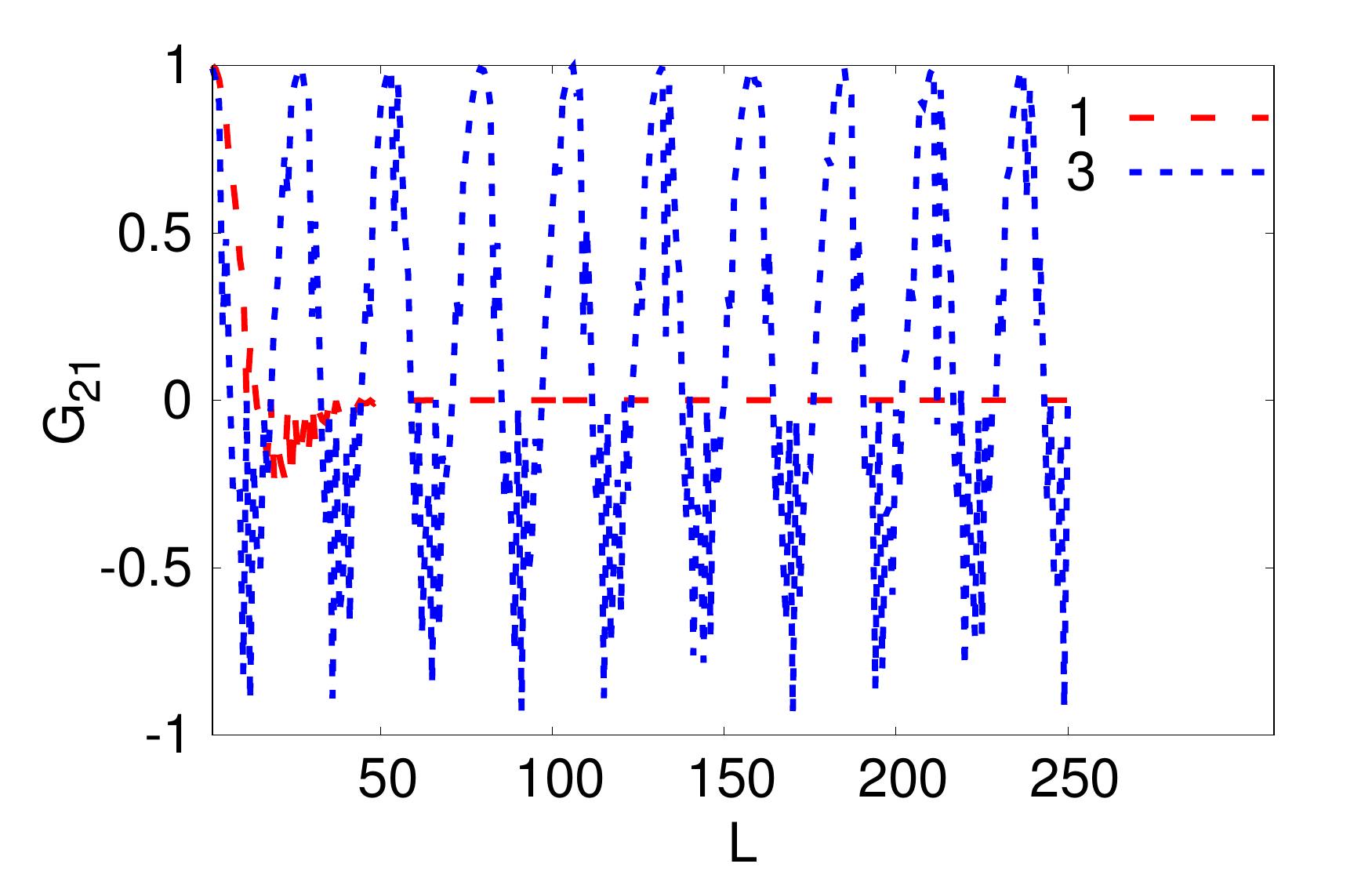} 
\caption{The zero bias transconductance as a function of length for the same choice
of parameters as in Fig.~\ref{fig:10} with the different values of $t'/\Delta$ shows the oscillatory behavior for $t'> 2\Delta$ .}
 \label{fig:11}
\end{figure}

The different behavior of transconductance for two different lengths
of the Kitaev ladder motivates a systematic study of its variation
with system size shown in Fig.~\ref{fig:10} and Fig.~\ref{fig:11}. Fig.~\ref{fig:10}(a)
or the dashed line in Fig.~\ref{fig:11} reveal that for $t^{\prime}<2\Delta$, the zero transconductance
region is dominant unless the system size is below a characteristic
length scale. The initial red region in Fig.~\ref{fig:10}(a) characterizes ET due to presence
of the decaying modes where the ladder length is so small that
an electron can tunnel through the ladder. This is a reflection of change
of the nature of transport~\cite{tan2015cooper,1367-2630-17-10-103016} from ballistic to diffusive as the length
is increased.

 For $t^{\prime}>2\Delta$ oscillatory behavior kicks in as
a result of the Fabry-P\'erot resonance phenomenon described
earlier. Therefore as shown in Fig.~\ref{fig:10}(b) and dotted lines in Fig.~\ref{fig:11}, both ET and CAR are
enhanced periodically.
 The region $|eV|<\Delta$ corresponds to all
eight $k$'s being real while the region $|eV|>\Delta$ corresponds to
only four $k$'s being real-valued while the other four $k$'s complex
valued. Thus, there are less modes leading to interference in the
region $|eV|>\Delta$ compared to the region $|eV|<\Delta$ and this
reflects in the richer interference pattern in the latter region.

\subsection{Variation of $t^{\prime}$, $t^{\prime\prime}$ and $\phi$}
\begin{figure}[h!]
\includegraphics[scale=0.28]{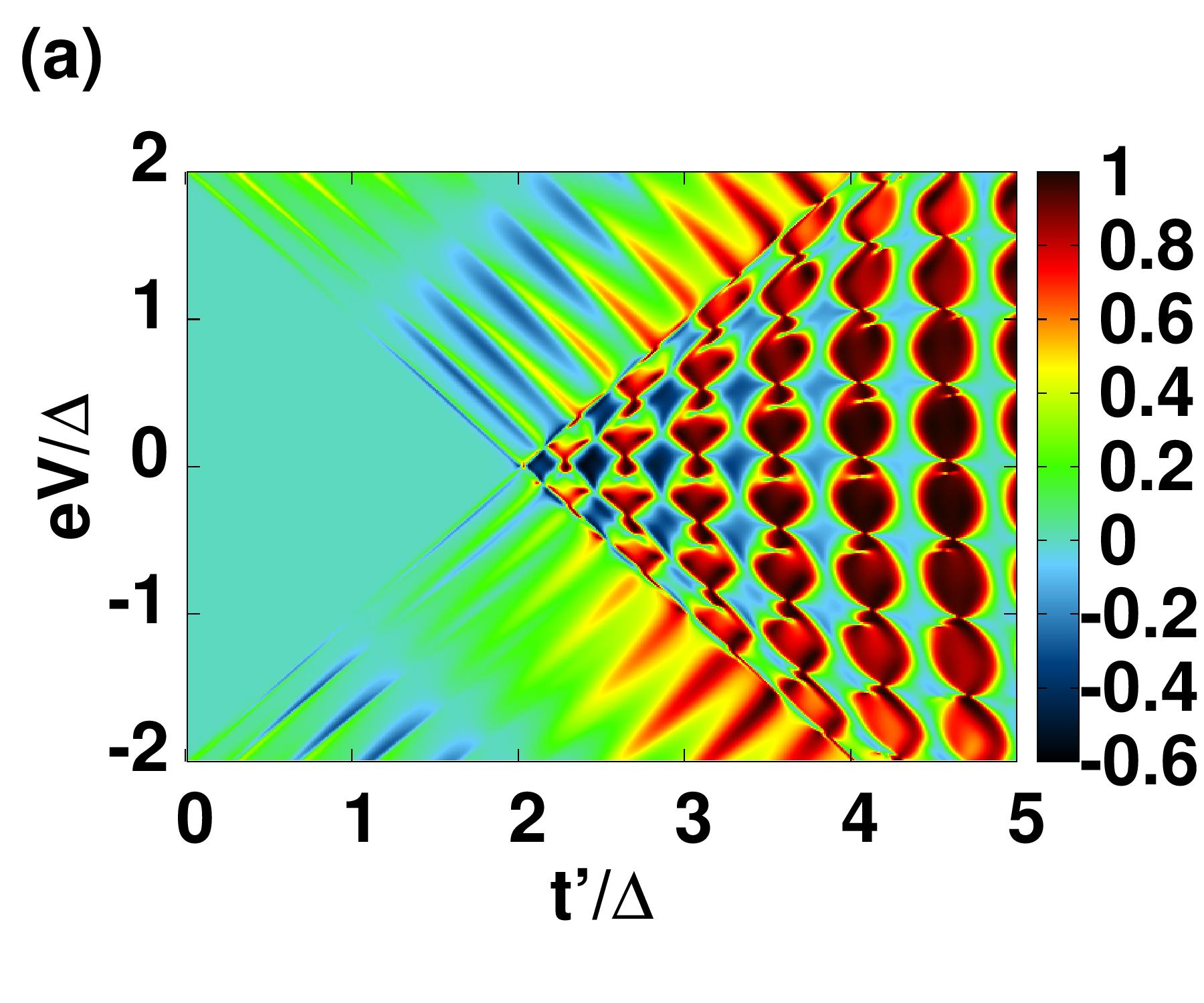}
\includegraphics[scale=0.28]{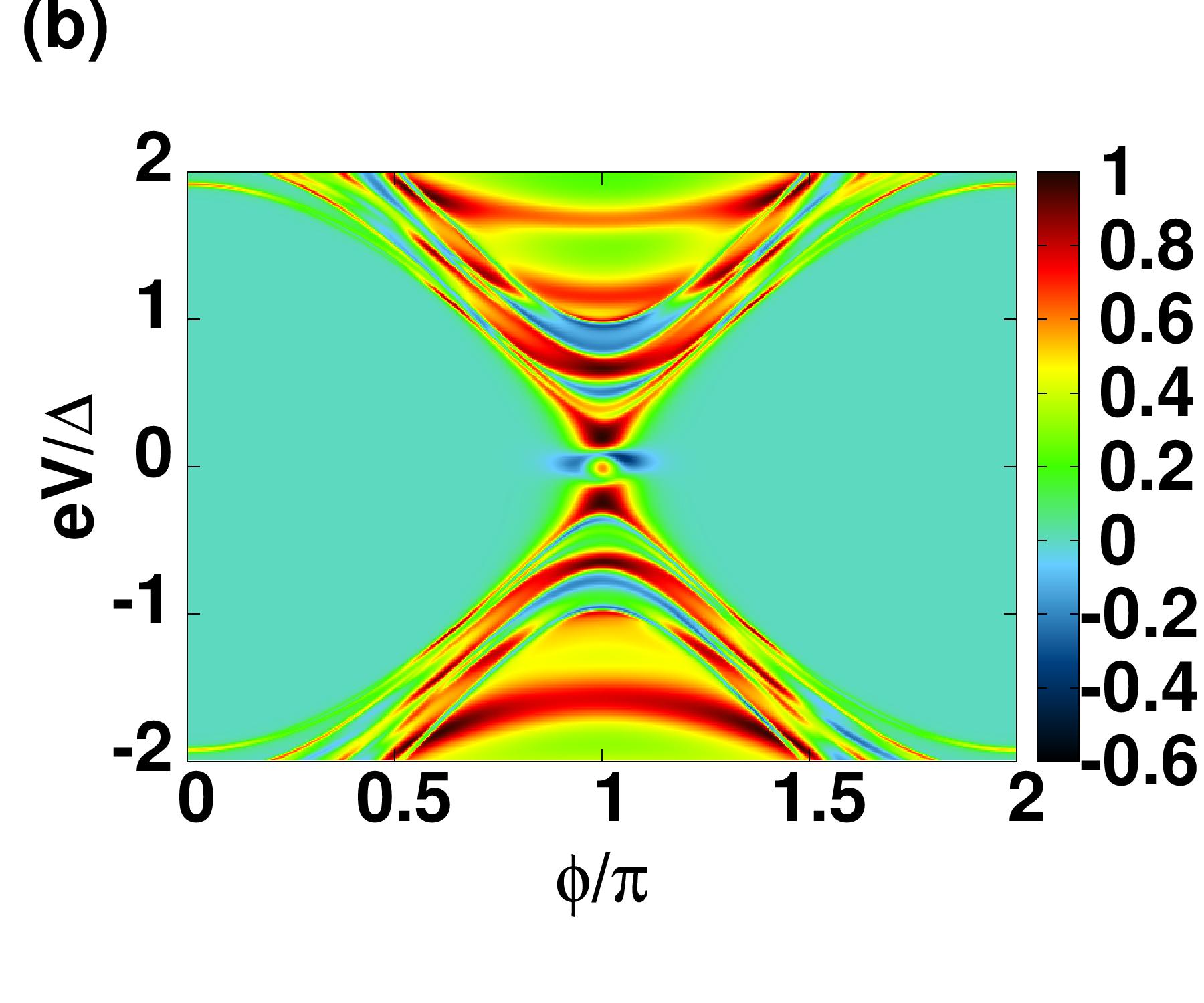}
\caption{The transconductance $G_{21}$ in units of $e^2/h$ as a function of bias and: (a)~inter-leg hopping of the ladder
for  $\phi=\pi$, (b)~phase difference for $t^{\prime}=3\Delta$.
The band crossing after a critical value of inter-leg hopping provides the propagating modes (a) and 
the enhancement is more prominent around  $\phi\approx\pi$ (b).  Parameters common to (a) and (b): 
$\Delta=0.1t$, $t''=t$, $\mu=0.50$, and $L=100$.}
\label{fig:12}
\end{figure}

Features of the transconductance plot as a function of bias $eV$ and the inter-leg hopping $t^{\prime}$ are presented in 
Fig.~\ref{fig:12}(a). At zero bias, for $t^{\prime}\le 2\Delta$, the transconductance is suppressed while for $t^{\prime}\ge 2\Delta$, the transconductance is enhanced periodically. At nonzero bias, there are three regions- the region where the transconductance is 
highly suppressed, the region where the transconductance is moderately enhanced and is periodic and the third region where the
transconductance is highly enhanced and periodic. These regions respectively  correspond to  all eight momenta in the ladder 
region imaginary, only four momenta in the ladder region real and all eight momenta in the ladder region real respectively. 

\begin{figure}[h!]
 \includegraphics[width=8cm]{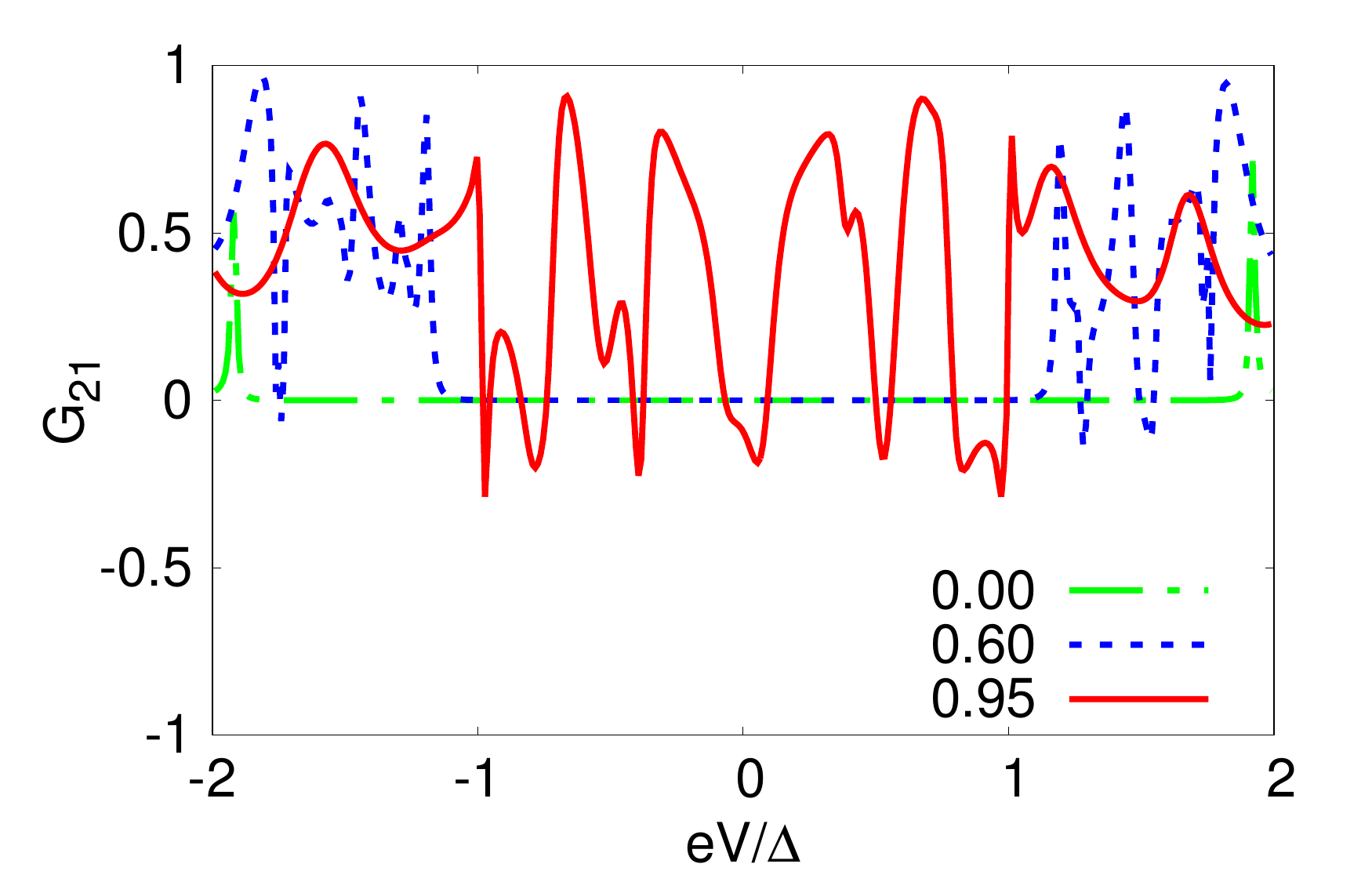}
 \caption{Transconductance in units of $e^2/h$ as a function of applied bias for the same choice of parameters as 
 in Fig.~\ref{fig:12}(b) with different values of $\phi/\pi$.}
 \label{fig:13}
\end{figure}

A substantial superconducting phase difference between the two legs of the
Kitaev ladder promotes enhanced transconductance. This can be seen
in Fig~\ref{fig:12}(b) and Fig.~\ref{fig:13}. 
Here, we have chosen the inter-leg hopping
$t^{\prime}=3\Delta$ so that there are subgap Andreev states in the
ladder for large phase difference $\phi$.  For small values of
the phase difference and values of the phase difference close to
$2\pi$, the transconductance is suppressed. For values of the
phase difference in the range $\pi/2\lesssim \phi\lesssim 3\pi/2$, we
see a rich interference pattern where CAR and ET are enhanced in
certain regions.  This interference pattern has origins in the
Fabry-P\'erot interference of the subgap Andreev states.
\begin{figure}[h!]
\includegraphics[scale=0.28]{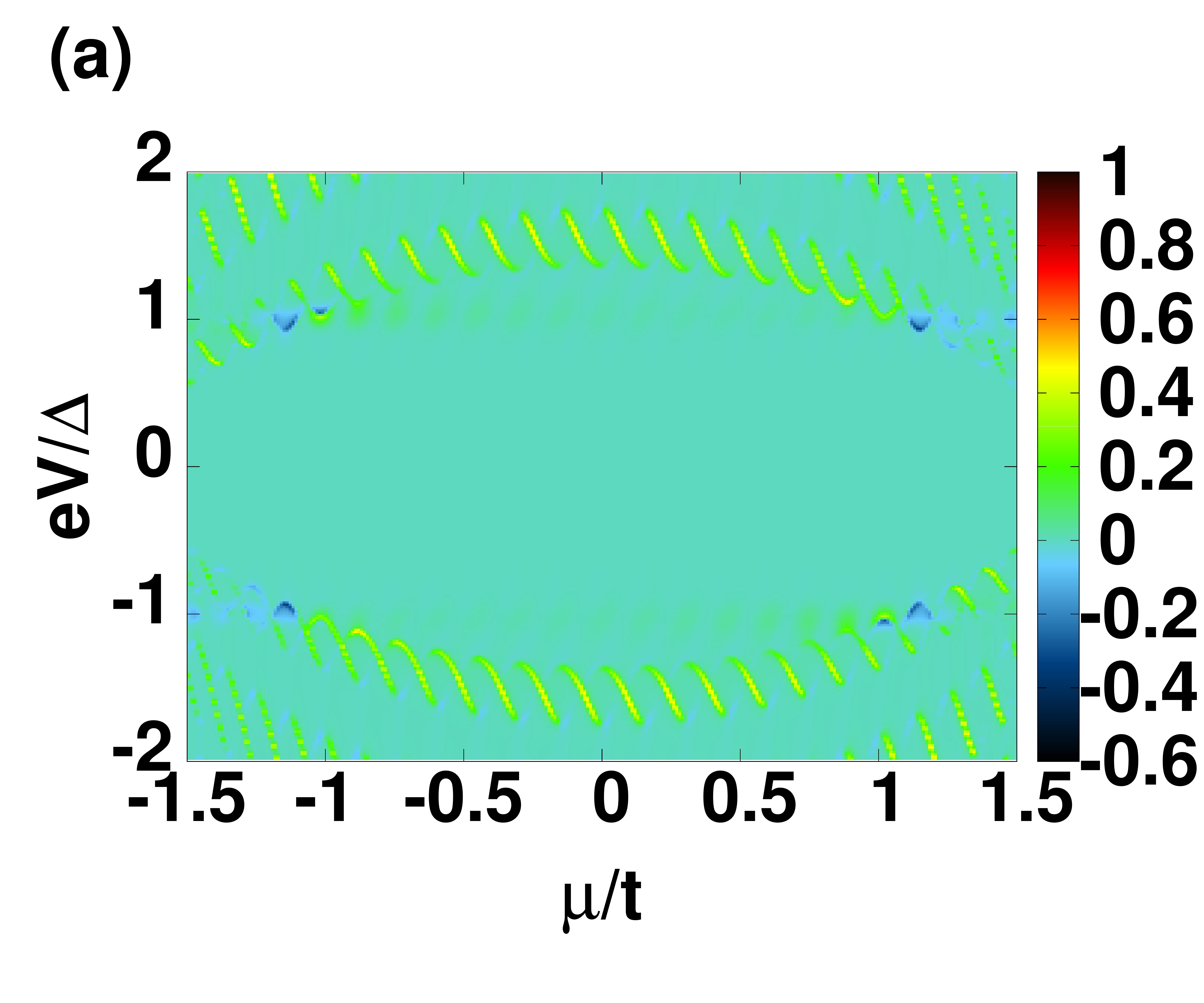}
\includegraphics[scale=0.28]{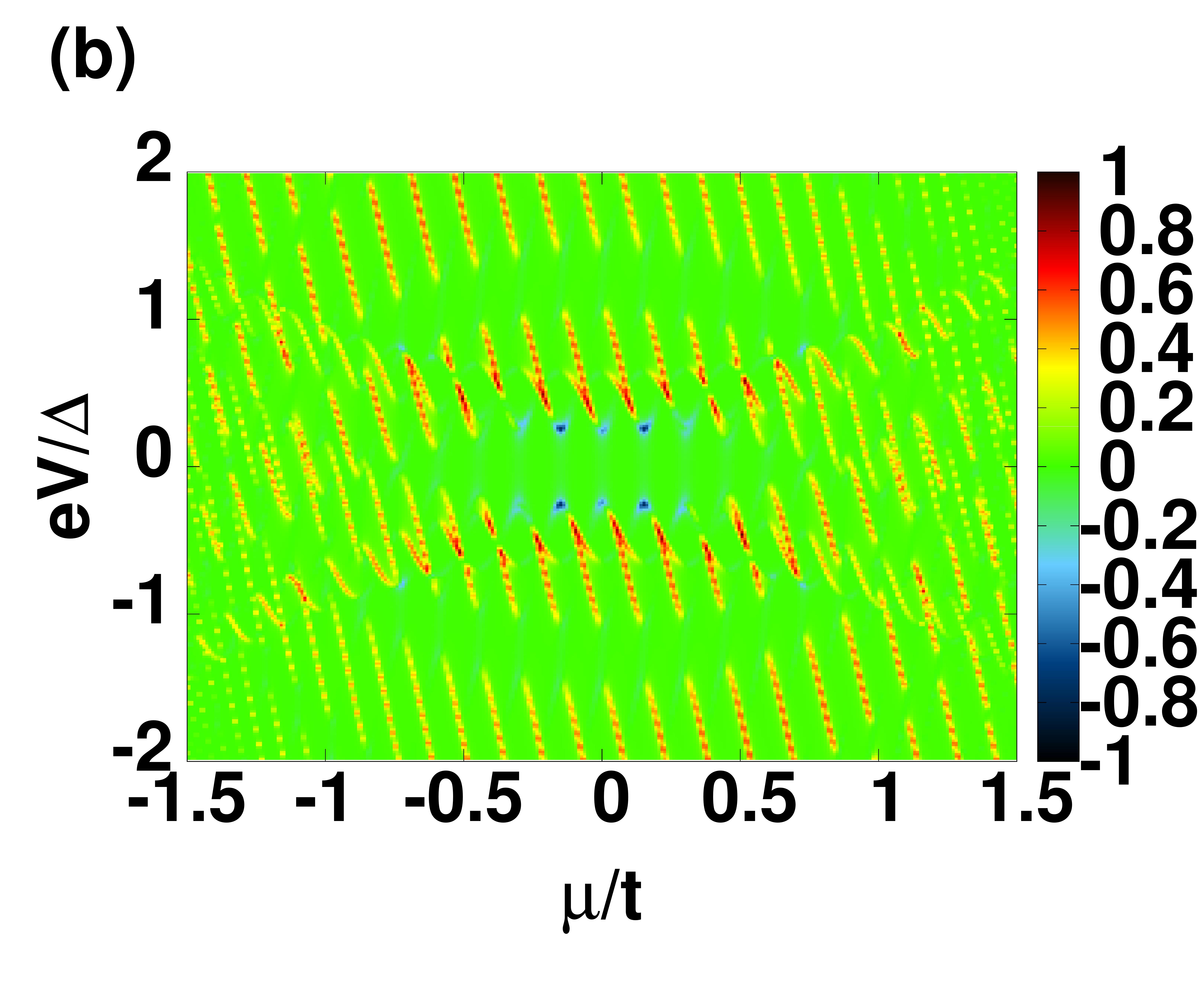}
\caption{The transconductance $G_{21}$ in units of $e^2/h$ for the parameters:
$\Delta=0.1t$, $t''=0.3t$, $\phi=\pi$, $L=40$ and (a) $t^{\prime}=\Delta$ (b) $t^{\prime}=3\Delta$.
The suppression of transport in most regions compared to Fig.~\ref{fig:4} can be seen .}
\label{fig:14}
\end{figure}
We have  studied a system where the junction is ideal and perfectly transmitting. In a
realistic system, the junction may not be perfectly transmitting. Such a junction can be 
modeled by changing the hopping term from normal metal lead to the SC ladder away from
hopping in the normal metal lead, that is $t''\neq t$. The ladder then is weakly connected to normal metal leads
and transport happens only at bias energies equal to  the energies of  the isolated ladder
hosting standing waves. The contour plot of the differential transconductance versus bias and chemical potential is
expected to  resemble more closely with Fig.~\ref{fig:8}. In Fig.~\ref{fig:14}, we plot a contour plot of 
the transconductance versus bias and chemical potential for the same choice of parameters as in Fig.~\ref{fig:4}, 
except that $t''=0.3t$ making the interface less transparent. This shows that it is important to have an interface 
which is as close to ideal as possible.
\section{Experimental Realization}
In this Section, we include a brief discussion of the possibility of an experimental realization of the proposed setup. 
While a Kitaev chain hosting Majorana 
fermions at its ends  is realized in semiconductor quantum wires~\cite{mourik2012signatures,albrecht2016exponential},
ladder systems have been discussed in recent literature~\cite{wakatsuki2014majorana}. Once two Kitaev chains are realized 
they can be separated by a thin gated insulator layer. The gate voltage can be used to tune the hopping between the two 
chains. A loop can be made between the superconductors proximetizing the two 
chains and a magnetic flux through such a loop can be used to control the superconducting phase difference as discussed in 
Ref.~\cite{keselman2013}. Thus, a Kitaev ladder with controllable phase difference and hopping can be realized. 
 We envisage that experimental groups will be motivated by our work to explore these 
directions further. 

Once the ladder is realized, another way to probe nonlocal transport is to measure current-current 
correlations~\cite{nilsson2008splitting}.
For this, both the normal metal leads are maintained at a bias $V$ and the ladder is grounded. Then equal amount of current 
flows in each of the leads. In this
configuration, the current that flows in the leads has contributions from local Andreev reflection and nonlocal CAR.
If $I_1(t)$ and $I_2(t)$ are currents at time $t$ in NM1 and NM2 respectively, the nonlocal noise power can be defined as 
$P_{12}=\int_{-\infty}^{\infty}dt \overline{\delta I_1(0)\delta I_2(t)}$ and the corresponding Fano factor is 
$F_{12}=P_{12}/e\bar{I_1}$, where $\bar{I_i}$ is time averaged current in lead $i$ and $\delta I_i(t) =  I_i(t) 
-  \bar{I_i}$ is the current fluctuation in lead $i$. A positive maximal value of $F_{12}=1$ corresponds to
maximally enhanced CAR.

\section{Summary} 
To summarize, we  studied a ladder consisting of two Kitaev chains maintained at a superconducting 
phase difference and connected to leads at either ends. We see that a nonzero phase difference and a sufficiently 
large inter-leg hopping 
generates plane wave states within the superconducting gap of the isolated ladder. We call these subgap Andreev states. 
The gap of the spectrum closes for sufficiently large inter-leg hopping which depends on the choice of the chemical potential
(for $\mu=0$, the gap closes when $t^{\prime}\geq2\Delta$) and for the 
choice  $\phi=\pi$.  We showed that the  subgap Andreev states   are responsible for enhanced crossed Andreev
reflection and enhanced electron tunneling. For a long ladder, one can see a resemblance in the energy spectrum 
of the isolated ladder and the differential transconductance indicating that the patterns in the differential
transconductance are due to the resonant levels present in the ladder region. We studied the dependence of the
transconductance on various parameters such as the bias, chemical potential, length of the ladder, inter-leg 
hopping strength and the phase difference. We find that by tuning the parameters, one can get values of negative 
transconductance with high  magnitude which indicates enhanced crossed Andreev reflection. We find periodic 
patterns in transconductance when the subgap Andreev states exist in the ladder and the 
periodic patterns can be understood as originating from the Fabry-P\'erot resonance between the plane wave modes
in the ladder. We also show that it is important to have a transparent interface between the ladder and the leads 
to enhance crossed Andreev reflection. 
\acknowledgements
A.S is grateful to SERB for the startup grant (File Number: YSS/2015/001696) and to DST-INSPIRE Faculty Award 
[DST/INSPIRE/04/2014/002461].
D.S.B acknowledges PhD fellowship support from UGC India.
A.~Soori thanks DST-INSPIRE Faculty Award (Faculty Reg. No.~:~IFA17-PH190)
for financial support, Subroto Mukerjee for discussions and Swathi Hegde for a
 discussion related to coding. This work started while 
A.~Soori was at IISER Bhopal. A.~Soori thanks IISER Bhopal for kind hospitality. 
\bibliography{ref1}
\onecolumngrid \appendix
\section{Equation of motion at the boundaries}
In this appendix, we have given a detailed calculation of the various scattering amplitudes. 
The plane wave solution for electrons and holes in the metallic regions are :
\begin{equation}\label{eqapp1}
\psi_{n,e}=\begin{cases}
\;e^{ik_ean}+r_e e^{-ik_ean}&{\rm for ~~ }n\leq0,\\
\;t_e e^{ik_ean}&{\rm for ~~}n\geq L+1,
\end{cases},
\qquad
\psi_{n,h}=\begin{cases}
\;r_he^{ik_han}&{\rm for~~}n\leq0,\\
\;t_h e^{-ik_han}&{\rm for ~~}n\geq L+1,
\end{cases},
\end{equation}
where, $r_e$, $r_h$ are reflection coefficients and $t_e$, $t_h$ are transmission coefficients of electrons and holes in the two metallic regions. 
Also, $k_e=\cos^{-1}\big(\frac{E+\mu}{2t}\big)$ and $k_h=\cos^{-1}\big(\frac{E-\mu}{2t}\big)$ are momenta of electron and hole respectively.

\begin{figure}[h!]
\center
\includegraphics[scale=0.88]{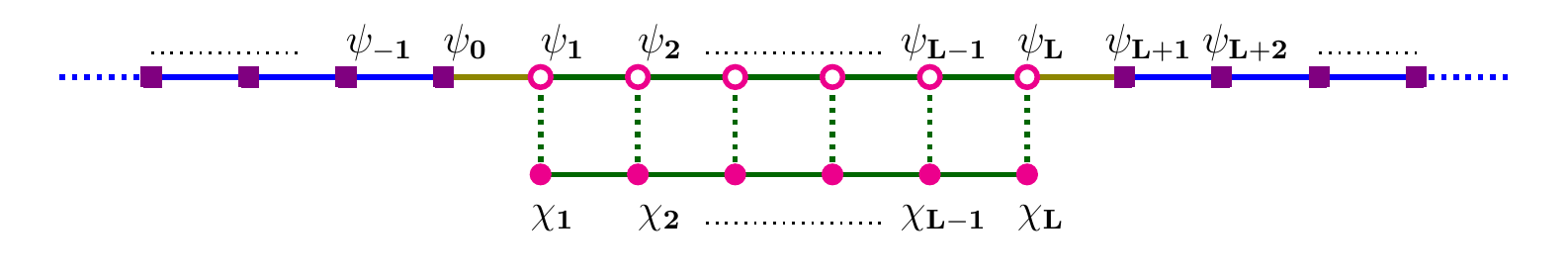}
\caption{A schematic of the system delineating the wavefunctions at the different sites of the system.}
\label{figappendix}
\end{figure}

The wavefunction for the ladder system is given by
\begin{equation}\label{eqapp2}
\psi_n=\displaystyle\sum_{\lambda,\nu,p}C_{\lambda,\nu,p}e^{i\lambda k_{\nu,p}an}
[\psi_{e,\lambda,\nu,p}~,~\psi_{h,\lambda,\nu,p}],
\end{equation}
\begin{equation}\label{eqapp3}
\chi_n=\displaystyle\sum_{\lambda,\nu,p}C_{\lambda,\nu,p}e^{i\lambda k_{\nu,p}an}
[\chi_{e,\lambda,\nu,p},\chi_{h,\lambda,\nu,p}]
\end{equation}
where, the different quantities are same as defined in the main text
(Section~\ref{section3}). For a given energy we have $8$ momenta
values  $\lambda k_{\nu,p}$'s (corresponding to $\pm$ values of the three variables
$\lambda,\nu,p$) which can be numerically found from
Eq.~\ref{eq:disp-ladder}. Overall there are $12$ unknowns as can be
seen from Eqs.~\ref{eqapp1},~\ref{eqapp2} and \ref{eqapp3}.  These
unknowns can be calculated by writing the equations of motion from the
Hamiltonian (Eq.~\ref{eq:fullham}) at the two boundaries as shown in
Fig~\ref{figappendix}.
These equations are:
\begin{align}
&E\psi_{0,e}=-t''\psi_{1,e}-t\psi_{-1,e}-\mu\psi_{0,e}\\
&E\psi_{0,h}=t''\psi_{1,h}+t\psi_{-1,h}+\mu\psi_{0,h}\\
&E\psi_{1,e}=-t\psi_{2,e}-t''\psi_{0,e}-\mu\psi_{1,e}-t^{\prime}\chi_{1,e}-\Delta e^{i\phi_1}\psi_{2,h}\\
&E\psi_{1,h}=t\psi_{2,h}+t''\psi_{0,h}+\mu\psi_{1,h}+t^{\prime}\chi_{1,h}+\Delta e^{-i\phi_1}\psi_{2,e}\\
&E\chi_{1,e}=-t\chi_{2,e}-\mu\chi_{1,e}-t^{\prime}\psi_{1,e}-\Delta e^{i\phi_2}\chi_{2,h}\\
&E\chi_{1,h}=t\chi_{2,h}+\mu\chi_{1,h}+t^{\prime}\psi_{1,h}+\Delta e^{-i\phi_2}\chi_{2,e}\\
&E\psi_{L+1,e}=-t\psi_{L+2,e}-t''\psi_{L,e}-\mu\psi_{L+1,e}\\
&E\psi_{L+1,h}=t\psi_{L+2,h}+t''\psi_{L,h}+\mu\psi_{L+1,h}\\
&E\psi_{L,e}=-t''\psi_{L+1,e}-t\psi_{L-1,e}-\mu\psi_{L,e}-t^{\prime}\chi_{L,e}+\Delta e^{i\phi_1}\psi_{L-1,h}\\
&E\psi_{L,h}=t''\psi_{L+1,h}+t\psi_{L-1,h}+\mu\psi_{L,h}+t^{\prime}\chi_{L,h}-\Delta e^{-i\phi_1}\psi_{L-1,e}\\
&E\chi_{L,e}=-t\chi_{L-1,e}-\mu\chi_{L,e}-t^{\prime}\psi_{L,e}+\Delta e^{i\phi_2}\chi_{L-1,h}\\
&E\chi_{L,h}=t\chi_{L-1,h}+\mu\chi_{L,h}+t^{\prime}\psi_{L,h}-\Delta e^{-i\phi_2}\chi_{L-1,e}.
\end{align}
Finally, the various unknowns can be calculated by solving these equations. The transconductance can be then calculated from Eq.~\ref{eq:13}.  
\end{document}